\documentclass{article}

\usepackage{arxiv}

\usepackage[utf8]{inputenc} % allow utf-8 input
\usepackage[T1]{fontenc}    % use 8-bit T1 fonts
\usepackage{hyperref}       % hyperlinks
\usepackage{url}            % simple URL typesetting
\usepackage{booktabs}       % professional-quality tables
\usepackage{amsfonts}       % blackboard math symbols
\usepackage{nicefrac}       % compact symbols for 1/2, etc.
\usepackage{microtype}      % microtypography
\usepackage{amsmath}
\usepackage{cleveref}       % smart cross-referencing
\usepackage{lipsum}         % Can be removed after putting your text content
\usepackage{xcolor}
\usepackage{subfigure}
\usepackage{url}
\usepackage{diagbox}
\usepackage[title]{appendix}
\usepackage[switch]{lineno}
\usepackage{graphicx}
\usepackage{tikz}
\usepackage{enumitem}
\usepackage{multirow}
\usepackage{algorithm} 
\usepackage{algorithmicx}
\usepackage{algpseudocode}
\usepackage{bbm}
\usepackage{wrapfig} 
\usepackage{caption}
\usepackage{ragged2e}

\newcommand{\yifan}[1]{{#1}}
\newcommand{\zhaoyuan}[1]{{#1}}
\newcommand{\revision}[1]{{#1}}
\newcommand{\eurosys}[1]{{#1}}
\newcommand{\revised}{}
\newcommand{\newrevised}{}

\long\def\comment#1{}

\newcommand{\ie}{{\em i.e.}}
\newcommand{\eg}{{\em e.g.}}
\newcommand{\etal}{{\em et al.}}

\newcommand{\para}[1]{\smallskip\noindent {\bf #1}}

\newcommand{\squishlist}{
\begin{list}{$\bullet$}
{ \setlength{\itemsep}{0pt}      \setlength{\parsep}{3pt}
        \setlength{\topsep}{3pt}       \setlength{\partopsep}{0pt}
        \setlength{\leftmargin}{3.5mm} \setlength{\labelwidth}{1em}
        \setlength{\labelsep}{0.5em} } }

\newcommand{\squishlisttwo}{
\begin{list}{-}
        { \setlength{\itemsep}{0pt}    \setlength{\parsep}{0pt}
                \setlength{\topsep}{0pt}     \setlength{\partopsep}{0pt}
                \setlength{\leftmargin}{1em} \setlength{\labelwidth}{1.5em}
                \setlength{\labelsep}{0.5em} } }

\newcommand{\squishend}{
\end{list}  }

\title{Enabling Real-time Neural Recovery for Cloud Gaming on mobile devices}

\date{}

\author{Zhaoyuan He\textsuperscript{1},~~~~
Yifan Yang\textsuperscript{2},~~~~
Shuozhe Li\textsuperscript{1},~~~~
Diyuan Dai\textsuperscript{1},~~~~ 
Lili Qiu\textsuperscript{1,2},~~~~
Yuqing Yang\textsuperscript{2}\\
\textsuperscript{1}Department of Computer Science, The University of Texas at Austin, Austin, TX, USA\\
\textsuperscript{2}Microsoft Research Asia Shanghai, Shanghai, P.R.China}

\renewcommand{\headeright}{}
\renewcommand{\undertitle}{}
\hypersetup{
pdftitle={A template for the arxiv style},
pdfsubject={q-bio.NC, q-bio.QM},
pdfauthor={David S.~Hippocampus, Elias D.~Striatum},
pdfkeywords={First keyword, Second keyword, More},
}

\begin{document}
\maketitle

\begin{abstract}
Cloud gaming is a multi-billion dollar industry. A client in cloud gaming sends its movement to the game server on the Internet, which renders and transmits the resulting video back. In order to provide a good gaming experience, a latency below 80 ms is required. This means that video rendering, encoding, transmission, decoding, and display have to finish within that time frame, which is especially challenging to achieve due to server overload, network congestion, and losses. In this paper, we propose a new method for recovering lost or corrupted video frames in cloud gaming. Unlike traditional video frame recovery, our approach uses game states to significantly enhance recovery accuracy and utilizes partially decoded frames to recover lost portions. We develop a holistic system that consists of (i) efficiently extracting game states, (ii) modifying H.264 video decoder to generate a mask to indicate which portions of video frames need recovery, and (iii) designing a novel neural network to recover either complete or partial video frames. Our approach is extensively evaluated using iPhone 12 and laptop implementations, and we demonstrate the utility of game states in the game video recovery and the effectiveness of our overall design.
\end{abstract}

% keywords can be removed
% \keywords{First keyword \and Second keyword \and More}

\vspace{-5pt}
\section{Introduction}
\label{sec:intro}
\vspace{-1pt}

\para{Motivation:} The cloud gaming market has been experiencing a rapid surge in popularity, with the global cloud gaming market size projected to grow at a CAGR of 64.1\% and reach \$14.01 billion by 2027~\cite{cloud-game-market}.
In cloud gaming, a player's input is transmitted to a remote game server on the Internet, which renders and sends back the resulting video for display on the client side. To provide a satisfactory gaming experience, high bandwidth and low latency are required. For instance, GeForce NOW requires at least 15Mbps for 720p (25Mbps for 1080p) and below 80ms latency (40ms for the best experience)~\cite{geforce_now_sys_req}. However, it is challenging to maintain a high throughput throughout the game session due to fluctuations in Internet performance and server load. If network congestion or server overload occurs, the video frames may not be delivered in time, leading to a disruption in the player's experience. 

\newrevised{While TCP retransmission can recover packet losses, its effectiveness is limited in cloud gaming due to the strict real-time requirements. Retransmissions are often too late to be useful, which is why most cloud gaming platforms opt for UDP for faster video transmission~\cite{di2021network}. Forward error correction (FEC) can  recover losses without adding extra delay, but it is challenging to predict the packet loss rate in advance and properly adjust the amount of FEC to add. Moreover, 
FEC is expensive: as reported in \cite{RFEC}, WebRTC increases the FEC to 50\% of original video data under 4\% packet loss, and to 80\% under 10\% packet loss. There are also various proposals to enhance video codecs, but it is challenging to deploy a new codec on a large scale. Therefore, to maximize our impact, we seek an approach that can recover losses without adding significant delay, overhead, or modifying video codecs.} 
% As a result, despite significant effort, it is still common to experience partial or complete video frame losses that cannot be recovered in time.} 

% the strigent real-time requirement in cloud gaming makes retransmission less effective (\eg, most retransmissions are too late to be useful). \newrevised{Therefore, most cloud gaming platforms use UDP for faster video transmission instead of TCP~\cite{di2021network}.} Forward error correction (FEC) is a common way to reduce loss rate without incurring extra delay. However, it is challenging to predict the packet loss rate in advance and properly adjust the amount of FEC to add. As a result, despite significant work in networking optimization and video codec, it is still common to have partial or complete video frame losses that cannot be recovered in time during cloud gaming. 

\begin{figure}[t!]
  \centering
  \includegraphics[width=0.7\columnwidth]{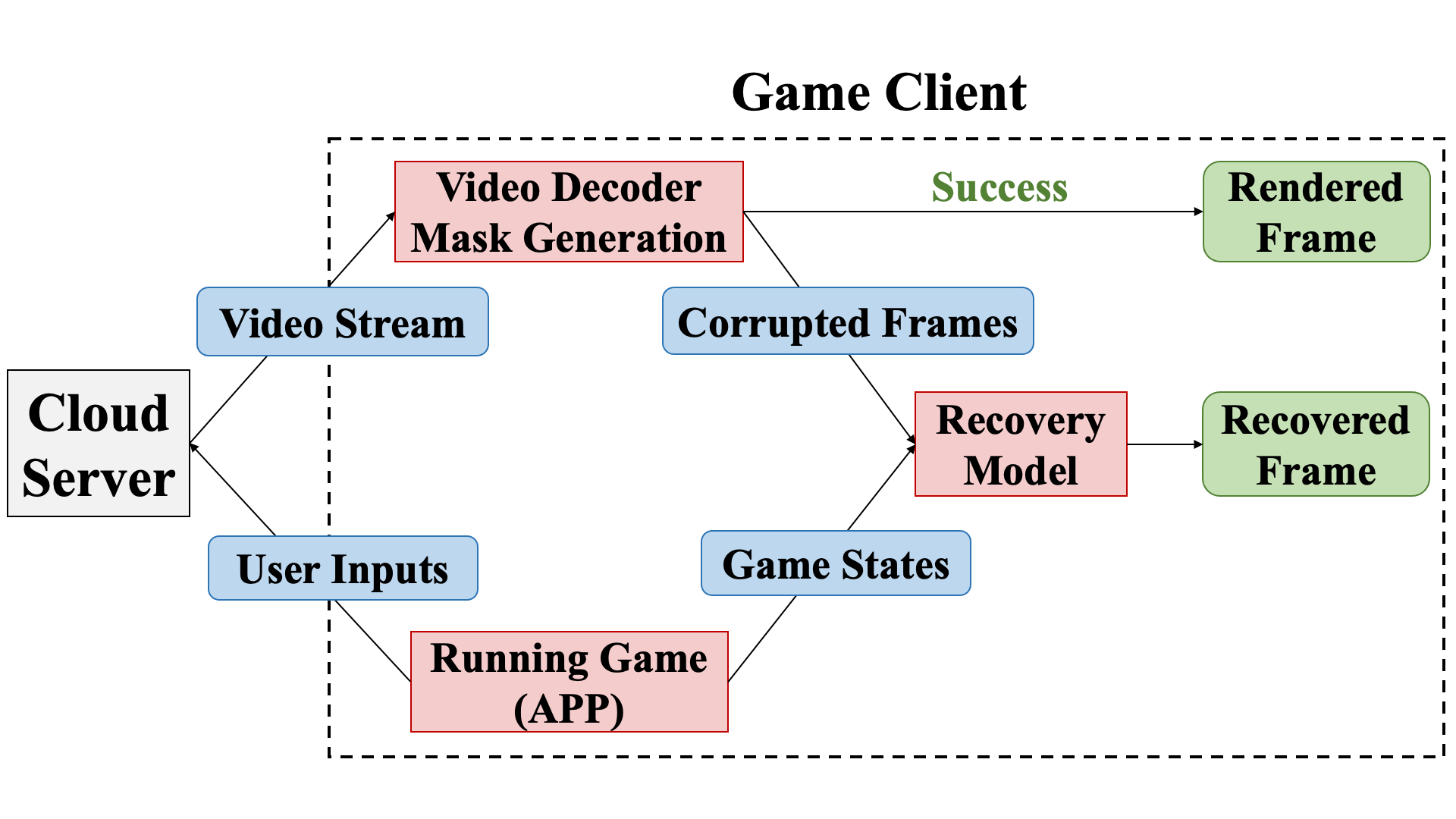}
  \vspace{-5pt}
  \caption{Overview of our video recovery system.}
  \label{fig:system_arch}
  \vspace{-5pt}
\end{figure}

\para{Our approach:} Motivated by the observation, we explore how to recover video frame(s) on the client side whenever the client does not receive them in time. 
% Interestingly, video prediction and video error concealment have attracted lots of work in general contexts (\eg, \cite{video-pred-survey,cnn1,rnn1,sdc-net,latent,sankisa2018video,sankisa2020temporal}). Unlike general methods, in cloud gaming the client has the game state information, which includes game objects' shapes, sizes, colors, orientations, positions, and movements. Such information is valuable for video recovery. 
Video prediction and error concealment have been extensively researched in various contexts (\eg, \cite{video-pred-survey,cnn1,rnn1,sdc-net,latent,sankisa2018video,sankisa2020temporal}), but in cloud gaming, the client possesses game state information that includes crucial details such as the shapes, sizes, colors, orientations, positions, and movements of game objects. This information can be highly beneficial for video recovery purposes.

To leverage game state information for video recovery, there are three critical research questions that need to be addressed: (i) how to efficiently extract and represent the game states, (ii) how to effectively utilize the game states for video recovery, and (iii) to what extent can game states assist in video recovery. 

To answer these questions, we develop a novel video recovery system shown in Figure~\ref{fig:system_arch}. 
\newrevised{First, the game client can efficiently extract game states from a running game app without rendering the whole frame. We only need to render foreground interactions and deformable objects.
% The game state can be either (i) extracted at the client side by running a game app on a mobile device or (ii) extracted at the server and transmitted reliably to the client. Both are practical.
% We implement (i) to only render foreground interaction and deformable objects. 
% which takes only XXX\% CPU and XXX\% GPU resources.  
We create a Compute Shader that transforms the 3D vertices of each visible game object into a 2D screen. To speed up game state extraction, we preload all objects' vertices into GPU at the beginning of games and create an array to indicate whether game objects exist in the current frame, thereby minimizing the overhead of uploading information from CPU to GPU. We further experiment with different game state representations, sizes, and resolutions to identify the appropriate configuration that balances the speed and accuracy. Note that the game state can also be extracted at the server and transmitted reliably to the client.}

% \newrevised{The game client has a running game application (APP) to efficiently generate game states when users play the game. Instead of rendering the whole game frame, the game APP only renders foreground interactions and deformable objects, thereby taking up only XX\% CPU and XX\% GPU usage. The game client also has a video decoder to generate a mask to indicate which portions of frames need recovery.}

% The game states and corrupted frames are fed to our recovery model for video recovery.} Our system explicitly designs each component as follows.} 

\revised{We then modify the H.264 video decoder using FFmpeg to generate a mask that indicates which parts of a video frame have decoding errors and require recovery. The game states and corrupted frames are fed into our recovery model for video recovery.}

% whether each $4\times4$ sub-macroblock of a 1080p frame can be correctly decoded. The mask can tell us which part of the frame is distorted such that our recovery model can further recover the lost part.}

Next, we design a novel neural network that can leverage game states to recover both complete and partial video frames. We first compute the optical flow between the current and previous game states and use it to warp the previous RGB frame to the current one. To improve the image quality, enhancement and inpainting modules are included to not only align pixels but also change colors and generate new content to better restore the image. 
\yifan{In this paper, we \revision{focus on a hybrid system that splits the rendering between clients and cloud servers.} We present a novel design that systematically implements our recovery model on both iPhone 12 and laptop devices, enabling the support of 30 FPS cloud gaming. Through extensive evaluations under diverse network conditions, we demonstrate significant improvements in recovered frames, achieving up to 9.7 dB PSNR. Furthermore, our approach effectively leverages game state information, resulting in a remarkable enhancement of up to 5.4 dB PSNR compared to schemes that do not utilize game states. To the best of our knowledge, this work is the first to exploit game state information for cloud game video recovery, showcasing a considerable boost in video quality for real-time mobile gaming. By demonstrating the feasibility of deep learning-based video frame recovery on mobile devices, our work paves the way for seamless and uninterrupted gaming experiences despite network losses. To optimize client resource usage, game states can be extracted on the server side and reliably transmitted using TCP, with minimal transmission overhead due to their compact size of 8 KB. We expect this work to inspire further advancements in efficiency, ultimately contributing to the improvement of cloud gaming performance on mobile devices.}
% \para{Paper outline:} The rest of the paper is organized as follow. In Section~\ref{sec:related}, we overview the related work. In Section~\ref{sec:extract},~\ref{sec:leverage}, and~\ref{sec:frame_recovery}, we describe our approach. In Section~\ref{sec:eval}, we systematically evaluate our approach. We conclude in Section~\ref{sec:conclusion}.

\vspace{-2pt}
\section{Related Work}
\label{sec:related}
\vspace{-2pt}

We classify existing work into the following three areas: (i) cloud gaming systems, (ii) video streaming, and (iii) video recovery.  

\para{Cloud gaming systems:} \cite{cai2016survey} provides a nice survey of cloud gaming research. There have been lots of works on optimizing cloud gaming systems, including virtual machine placement~\cite{VM-placement, han2020virtual}, distributed game engines~\cite{bulman2020cloud}, GPU sharing~\cite{GPU-sharing,GPU-sharing2}, cloud scheduling~\cite{cloud-schedule}, and resource allocation~\cite{marzolla2012dynamic,res-alloc2,res-alloc3}. Some works~\cite{fu2016models, wu2015enabling, wu2016streaming, alhilal2022nebula} leverage forward error correction (FEC) to % cancel out error propagation at the cost of a motion to photon latency while 
protect game frames from
losses, ~\newrevised{but FEC incurs significant overhead~\cite{RFEC}, which may make more frames miss their 33ms deadline for 30FPS games under a highly fluctuating network.} Outatime~\cite{lee2015outatime} renders speculative frames one RTT in advance.
% leading to 120ms of network latency. 
Furion~\cite{lai2017furion} employs a cooperative renderer architecture that renders foreground interactions locally on the phone while prefetching pre-rendered frames from servers to reduce delay. Game engine information has been used to speed up the motion estimation in video encoding H.264/AVC~\cite{ME}. 
% Mohammadi \etal~\cite{mohammadi2015object} develop a framework that encodes each object independently, controls its quality, and streams it using MPEG 4’s BiFS to the client. 
Chen \etal~\cite{chen2019framework} use collaborative rendering, progressive meshes, and 3D image warping to dynamically adapt image quality. 

% interaction delay with a tradeoff in image quality. 

\para{Video streaming:} To accommodate fluctuation in network performance, existing video streaming systems implement dynamic video rate adaptation. Many rate adaptation schemes have been proposed (\eg, QDASH~\cite{mok2012qdash}, FESTIVE~\cite{jiang2012improving}, client-side
buffer-based adaptation~\cite{tian2012towards}, \cite{houdaille2012shaping} and QAVA~\cite{chen2012qava}). \cite{huang2014buffer} adapts the rate based on buffer occupancy. \cite{yin2015control} develops a model predictive control (MPC) based on buffer occupancy and throughput estimation. \cite{huang2018qarc} develops rate adaptation based on deep reinforcement learning. To enhance the performance, several works improve the throughput prediction by exploiting correlation between video sessions (\eg, CFA~\cite{jiang2016cfa} and CS2P~\cite{sun2016cs2p}) or using ML approaches (\eg, \cite{pred-thesis,icc2020-pred}). 

There has been a lot of work on video delivery over LTE (\eg, \cite{4G-PERCEIVE,4G-VR}). Due to the recent deployment of 5G, video streaming over 5G has attracted increasing interest. 5G poses more challenges due to its larger fluctuation and unpredictability. \cite{5G-all} empirically compares several adaptive bit rate algorithms in 5G networks and reports 3.7\% to 259.5\% higher stall time than 4G/LTE. 

\zhaoyuan{Neural-enhanced video streaming has been well studied (e.g. NAS~\cite{yeo2018neural}, LiveNAS~\cite{kim2020neural}, NEMO~\cite{yeo2020nemo}, NeuroScaler~\cite{yeo2022neuroscaler}). These approaches employ super-resolution models to enhance low-resolution frames to high-resolution either on clients or media servers, aiming to enhance transmission efficiency. Different from these works, our work specifically addresses video recovery for cloud gaming, which presents a greater challenge due to the stringent real-time demands typically associated with gaming. In contrast to merely enhancing existing frames, video recovery in gaming involves generating new content instead of enhancing individual pixels.}

\para{Video recovery:} We classify the work of video recovery into two areas: video prediction and video error concealment. Many interesting algorithms have been proposed for video prediction. Several deep learning architectures have been used for video prediction, such as convolutional networks (\eg, \cite{cnn1,cnn2,cnn3,cnn4}), recurrent networks (\eg, \cite{rnn1,rnn2}), and generative networks. Among them, GANs are particularly effective for generation tasks. However, GANs are typically designed to generate realistic images, which may not be necessarily similar to the next frame. \cite{sdc-net} also learns transformations, such as pixel motion, to synthesize the next video frame from the previous ones. 
% \cite{latent} decouples motion from visual content to facilitate prediction. 

% comment for brevity
\comment{
The design of loss functions for video prediction is well explored. While pixel-wise losses, such as Cross Entropy, $l_1$, $l_2$, and Mean-Squared Error, are widely used, simply minimizing the losses may make the predictor average out multiple possible future frames and generate blurred images. To address the issue, researchers have explored adversarial training, probabilistic alternatives, and designing regularization terms to promote image sharpness (\eg, gradient difference loss~\cite{loss4}) and visual quality (\eg, perception loss~\cite{loss2,loss3}).}

% Therefore, we adapt the loss function used in GAN to make it generate image closer to the next frame while being realistic. [XXX: revisit the last sentence]

% (\eg, density based~\cite{density1,density2,density3}, and auto-encoder based \cite{autoencoder1,autoencoder2}). Amon

% Pioneers in the field have created numerous methods to approach the problem. Researchers such as XXXXXX start with predicting the raw pixel of the future frame by analysing the previous frame’s pixel-level details and scene dynamics. But, due to the unstableness and highly dimensional nature of the pixel space, prediction error can grow exponentially on the long-term horizon and the problem of extracting a robust representation from raw pixels is extremely complicated. As a result, researchers start to decompose the factors of variation from the frames.

\revised{Several other works utilize neural networks to realize video error concealment such that corrupted frames can be recovered from previous frames.~\cite{sankisa2018video} predicts optical flow from generated flows of past frames to reconstruct the degraded portion of the frame.~\cite{sankisa2020temporal} designs a capsule network framework to extract the decoded temporal dependencies, which are further combined with past frames and passed through a reconstruction network to perform motion-compensated error concealment.} 

\para{Summary:} Our work differs from the existing work as follows: (i) our work is the first one that extracts and leverages the game state for complete or partial video frame recovery and demonstrates significant benefits of game states, and (ii) unlike the existing work, which focuses only on the video reconstruction accuracy, our system accurately recovers videos in real-time on mobile devices.  
\comment{
\section{Background}
\label{sec:background}

% https://forum.unity.com/threads/trying-to-understand-the-rendering-steps.640438/
% https://medium.com/shader-coding-in-unity-from-a-to-z/rendering-pipe-line-f0471aa0904b
% https://docs.unity3d.com/Manual/render-pipelines-overview.html

Unity~\cite{unity} is a cross-platform game engine developed by Unity Technologies. It is primarily used to develop video games and simulations. At a high level, the Unity rendering pipeline consists of culling, rendering, and post-processing~\cite{unity-doc}. % Unity provides a few pre-built render pipelines and also allows users to create their own pipeline. 

For each frame, a camera performs frustum culling to remove renderers that fall outside the camera's view. Occlusion culling further removes the objects that are occluded. Both frustum culling and occlusion culling remove the need of rendering unnecessary objects, and enhance efficiency. 

The rendering step involves creating geometry, adding lighting effects, projecting the 3D environment onto the 2D screen based on the camera configuration. 

Unity offers several post-processing effects, such as blurring the background, blurring the object in motion, balancing colors around the white area, adding fog effects etc. These effects can be used to create specialized visual effects. 
}

\section{Motivation}
\label{sec:local-rendering}

\begin{figure}[ht!]
\centering
% \vspace{-10pt}
\subfigure[Viking Village]{%
\includegraphics[width=0.23\textwidth]{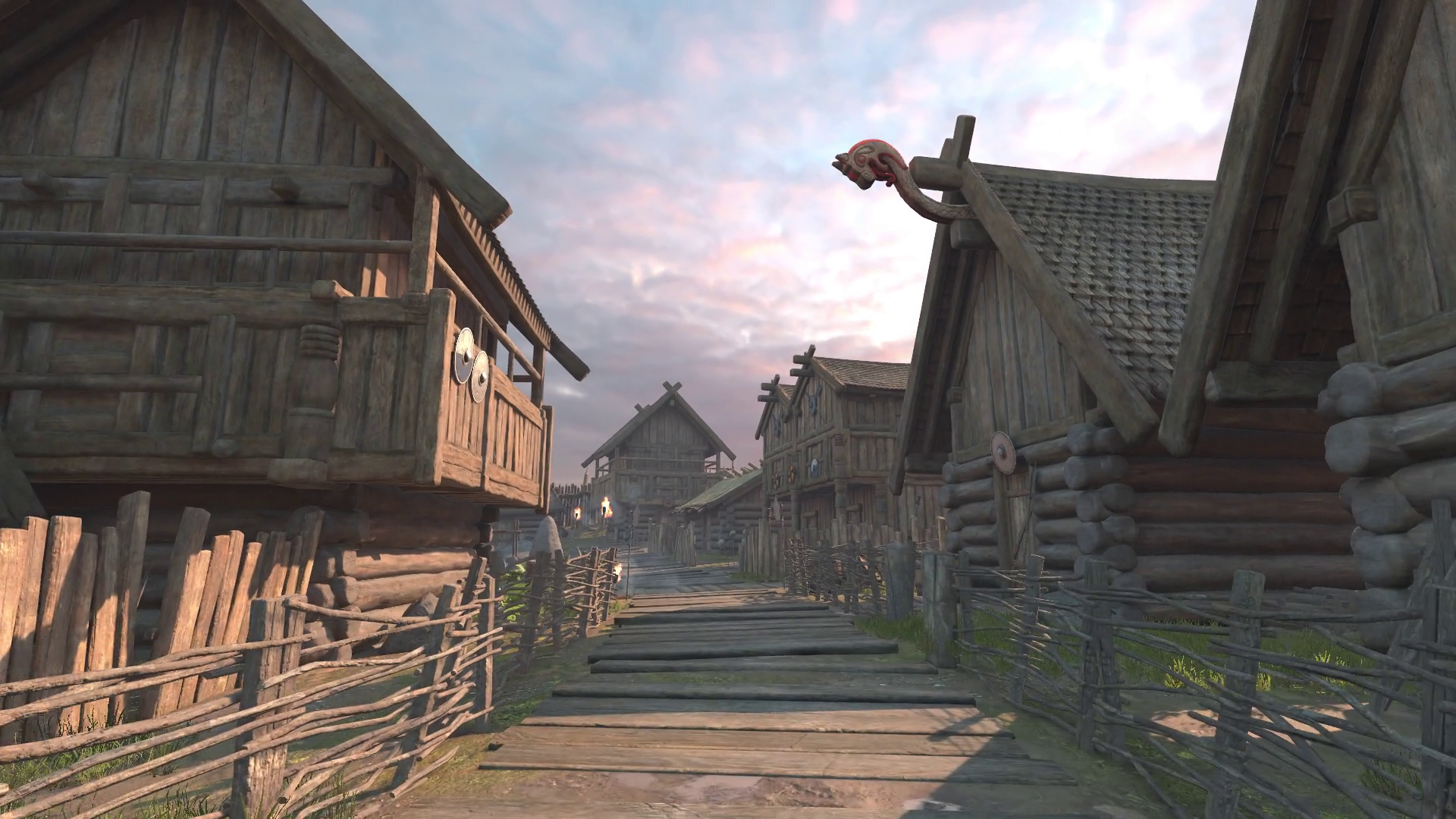}
\label{fig:vk}
}
\subfigure[Nature]{%
\includegraphics[width=0.23\textwidth]{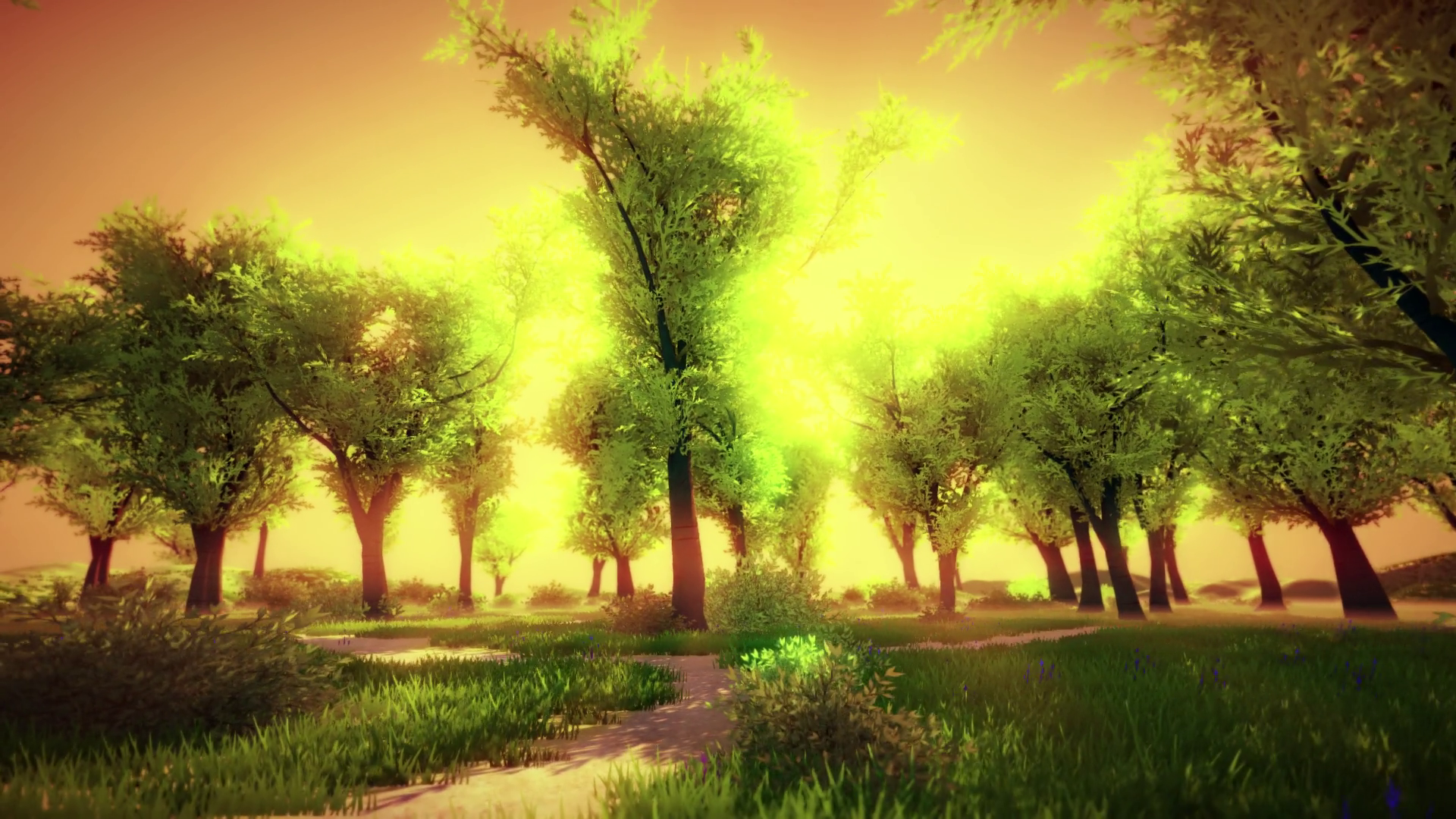}
\label{fig:nature}
}
\subfigure[Corridor]{%
\includegraphics[width=0.23\textwidth]{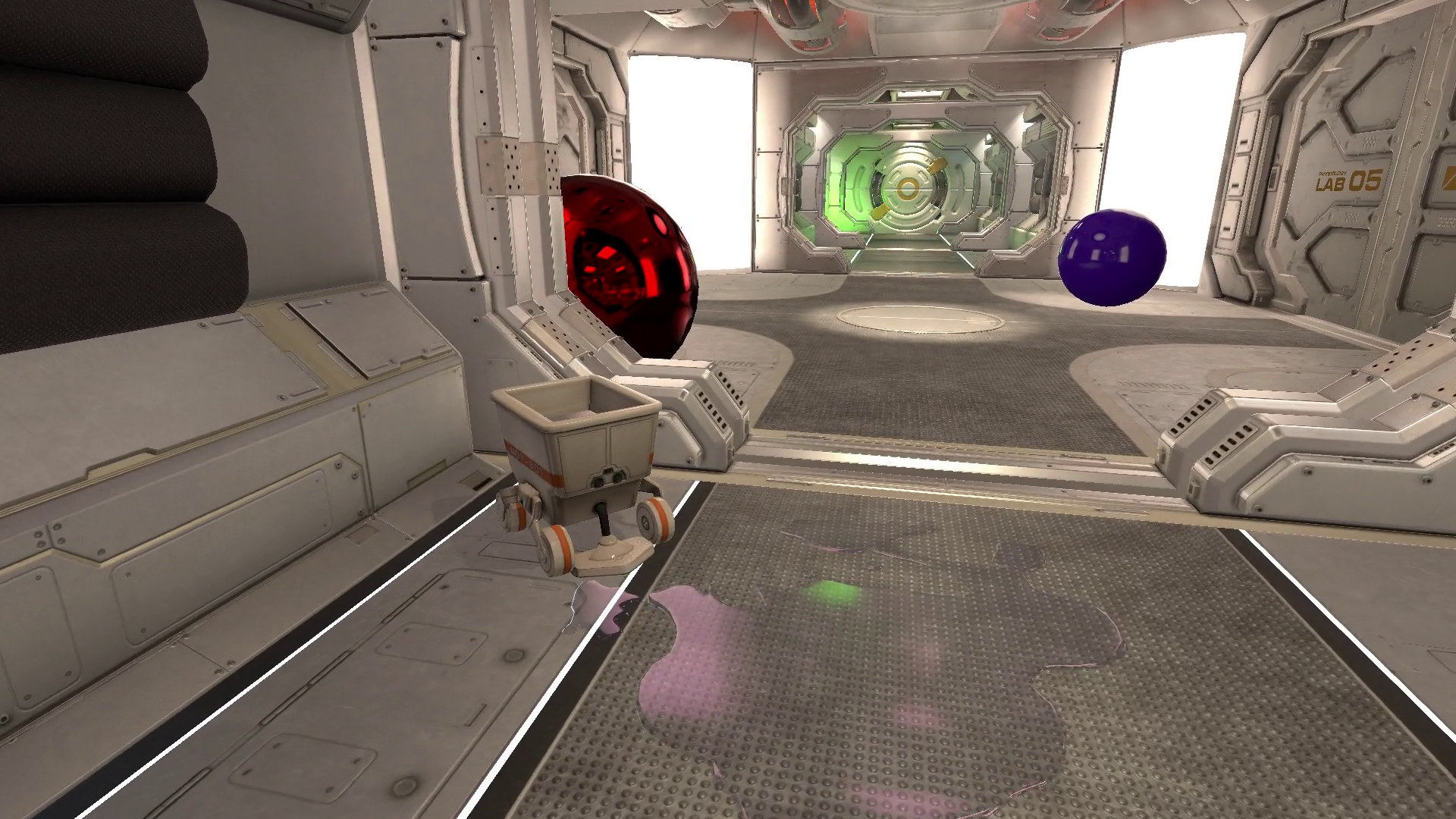}
\label{fig:corridor}
}
\subfigure[Viking Village+]{%
\includegraphics[width=0.23\textwidth]{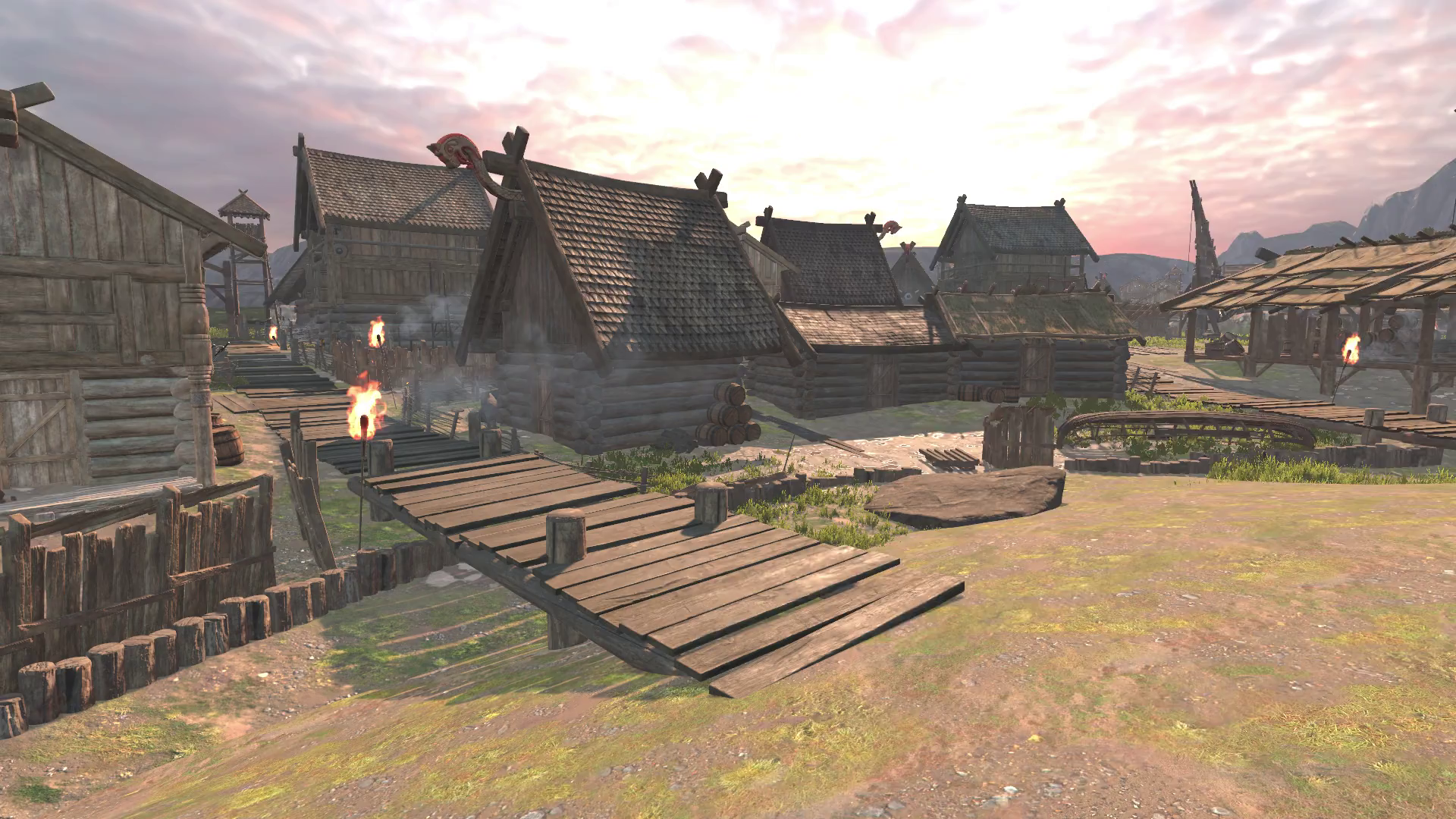}
\label{fig:vk+}
}
% \vspace{-10pt}
\caption{High-quality Unity games.}
\label{fig:games}
\end{figure}

\begin{figure}[H]  
  \centering  
  \begin{minipage}{0.48\textwidth}  
    \centering  
    \includegraphics[width=0.9\textwidth]{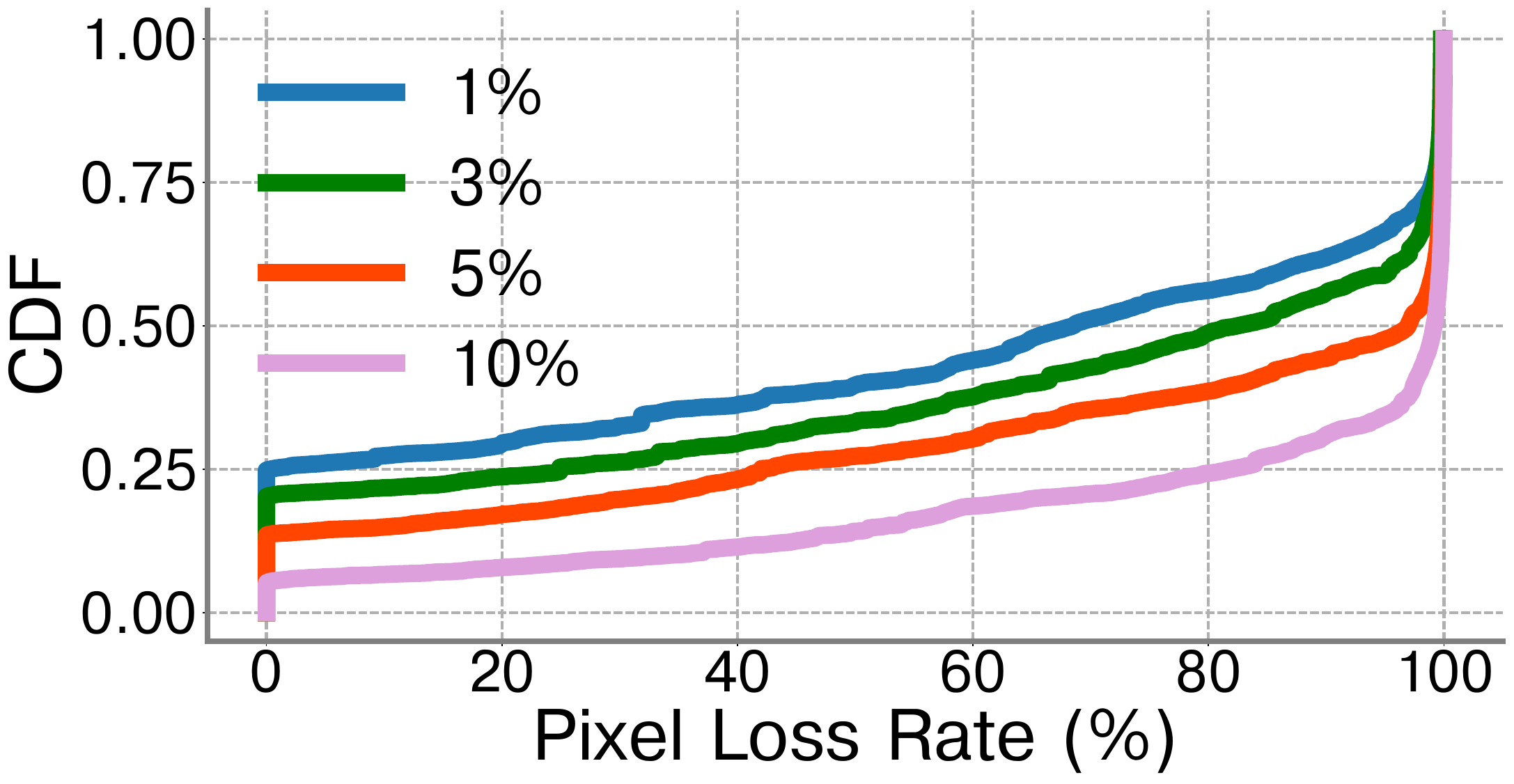}  
    % \vspace{-10pt}
    \caption{Average pixel loss rate of each frame under different network loss rates.}  
    \label{fig:partialloss} 
    % \vspace{-10pt}  
  \end{minipage}%  
  % \hfill  
  \begin{minipage}{0.48\textwidth}  
    \centering  
    % \vspace{-5pt}   
    % \captionof{table}{Frame rate when running different games}  
    \resizebox{0.72\columnwidth}{!}{%  
      \begin{tabular}{ccc}  
        \toprule  
        Machine & Game Name & FPS \\  
        \midrule  
        \multirow{3}{*}{iPhone 12} & Viking Village & 20  \\  
        & Nature & 23 \\  
        & Corridor & 17 \\ \hline  
        \multirow{1}{*}{Laptop} & Viking Village+ & 18  \\  
        \bottomrule  
      \end{tabular}   
    }
    \captionsetup{type=table}
    \caption{Frame rate when running different games}
    \label{tab:fps}
  \end{minipage}  
\end{figure}

% \begin{figure*}[t!]
%   \centering
%   \subfigure[Viking Village]{%
%     \includegraphics[width=0.46\columnwidth]{figures/vk.png}
%     \vspace{-20pt}
%     \label{fig:vk}
%   }
%   \subfigure[Nature]{%
%     \includegraphics[width=0.46\columnwidth]{figures/nature.png}
%     \vspace{-20pt}
%     \label{fig:nature}
%   }
%   \subfigure[Corridor]{%
%     \includegraphics[width=0.46\columnwidth]{figures/corridor.png}
%     \vspace{-20pt}
%     \label{fig:corridor}
%   }
%   \subfigure[Viking Village+]{%
%     \includegraphics[width=0.46\columnwidth]{figures/vk+.png}
%     \vspace{-20pt}
%     \label{fig:vk+}
%   }
%   \vspace{-12pt}
%   \caption{High-quality Unity games.}
%   \vspace{-10pt}
%   \label{fig:games}
% \end{figure*}

% In this section, we describe several observations that motivate our work. 
% We show that it is challenging to perform local rendering or video prediction in real time. Moreover, the current game video prediction accuracy is insufficient. According to our network analysis of an existing cloud gaming platform, we demonstrate that clients usually have idle cycles for video recovery.

\subsection{Local Rendering}

Network performance fluctuates widely~\cite{4G-VR,4G-PERCEIVE,5G-all}. When the network condition degrades, a natural solution is to perform local rendering on the client side. To assess the feasibility of this approach, we measure the resource usage and resulting frame rate when rendering on the two following local devices: (i) iPhone 12: Apple A14 Bionic Chipset, Hexa-core (2x3.1 GHz Firestorm + 4x1.8 GHz Icestorm) CPU and Apple GPU (4-core graphics), and (ii) laptop: Intel(R) Core(TM) i5-9300H at 2.40GHz CPU and GeForce GTX 1650 (Mobile) GPU. \zhaoyuan{With over 94\% of U.S. iPhone purchases in Q1 2023 being iPhone 12 and higher versions~\cite{CIRP}, it can be inferred that most users' smartphones should match or exceed the performance of the iPhone 12.} We consider the following three games in our measurements: Viking Village~\cite{vk}, Corridor~\cite{corridor} and Nature~\cite{nature}. They are high-quality virtual-world Unity apps with complex and realistic environments as shown in Figure~\ref{fig:games}. We use them for our experiments because they are originally designed for high-end PCs.

Table~\ref{tab:fps} shows that the first three games achieve only 17-23 frames per second (FPS) on the iPhone 12, which is insufficient to support a typical requirement of 30 FPS. The laptop has a more powerful CPU and GPU, and can achieve over 30FPS for these three games. However, as the games get more complex, local rendering can also become infeasible even for the laptop. For validation, we build a larger game based on Viking Village by increasing the game area and adding more game objects to the scene. We call the larger game Viking Village+, as shown in Figure~\ref{fig:vk+}. The laptop can only achieve 18 FPS for Viking Village+.   

These results show that existing local rendering is often infeasible and motivates us to seek alternatives to recover corrupted frames due to prolonged network or server delay. 

\subsection{Video Prediction}

Video prediction has been well studied and can help us recover lost frames. Many state-of-art video prediction models are too heavy to perform inference on the laptop in real-time (\eg, within $33ms$). For example, CrevNet~\cite{Yu2020Efficient} and PreRNN+~\cite{wang2021predrnn} take $600ms$ and $3s$ to predict 1080p frames on the laptop, respectively. 

An alternate way to predict frames is to calculate the optical flow between the last two frames and warp the previous frame using the optical flow. Unfortunately, state-of-art optical flow models are not fast enough to perform real-time inference for 1080p videos. For example, it takes SPyNet~\cite{ranjan2017optical} $250ms$ to compute the optical flow between two consecutive frames on the laptop. % Therefore, this situation makes our task quite challenging. 

% \subsection{Accuracy of Game Video Prediction}

It is challenging to accurately predict video frames due to uncertainties in future events. For example, CrevNet~\cite{Yu2020Efficient} achieves 0.49 SSIM and 19.2dB PSNR for game video prediction. Such accuracy is insufficient to support a satisfactory gaming experience. Meanwhile, we observe that video frames from a game are rendered based on the game states. This gives us an opportunity to leverage the game states to significantly improve the accuracy of video prediction.  

\subsection{Partial Recovery}

% \begin{figure}[t!]
%   \centering
%   \includegraphics[width=0.65\columnwidth]{figures/partial_pixel_loss_rate.pdf}
%   \vspace{-10pt}
%   \caption{Pixel loss rate of each frame when a fraction of the frame cannot be decoded. We add cross-traffic to the network to incur 1\%, 3\%, 5\%, and 10\% average packet loss rate.}
%   \label{fig:partialloss}
%   \vspace{-15pt}
% \end{figure}

\revised{To understand how packet losses affect the video frame delivery, we conduct the following measurement. We set up a video streaming session and add cross traffic to the network to incur around 1\%, 3\%, 5\%, and 10\% average packet loss rates. Figure~\ref{fig:partialloss} plots the fraction of corrupted pixels in video frames under various packet loss rates. As we can see, % 50\% corrupted video frames have within 90\% lost pixels after video decoding, and 
% 25\%, XXX\%, XXX\%, and XXX\% video frames are completely lost after decoding under the 1\%, 3\%, 5\%, and 10\% packet loss rates, respectively. 
25\% frames are completely lost and 50\% frames are partially corrupted after decoding under the 1\% packet loss rate. The numbers are even higher under a higher loss rate. Note that when reference frames are lost, all frames that depend on them are lost, as well. These results suggest that it is necessary to develop a video recovery approach that can recover both partial frame loss and complete frame loss.}

\begin{figure*}
\centering
\begin{minipage}[ht]{1.05\columnwidth}
\begin{minipage}{0.45\columnwidth}
    \begin{figure}[H]
        \centering
        % \vspace{-18pt}
        \includegraphics[width=0.75\columnwidth]{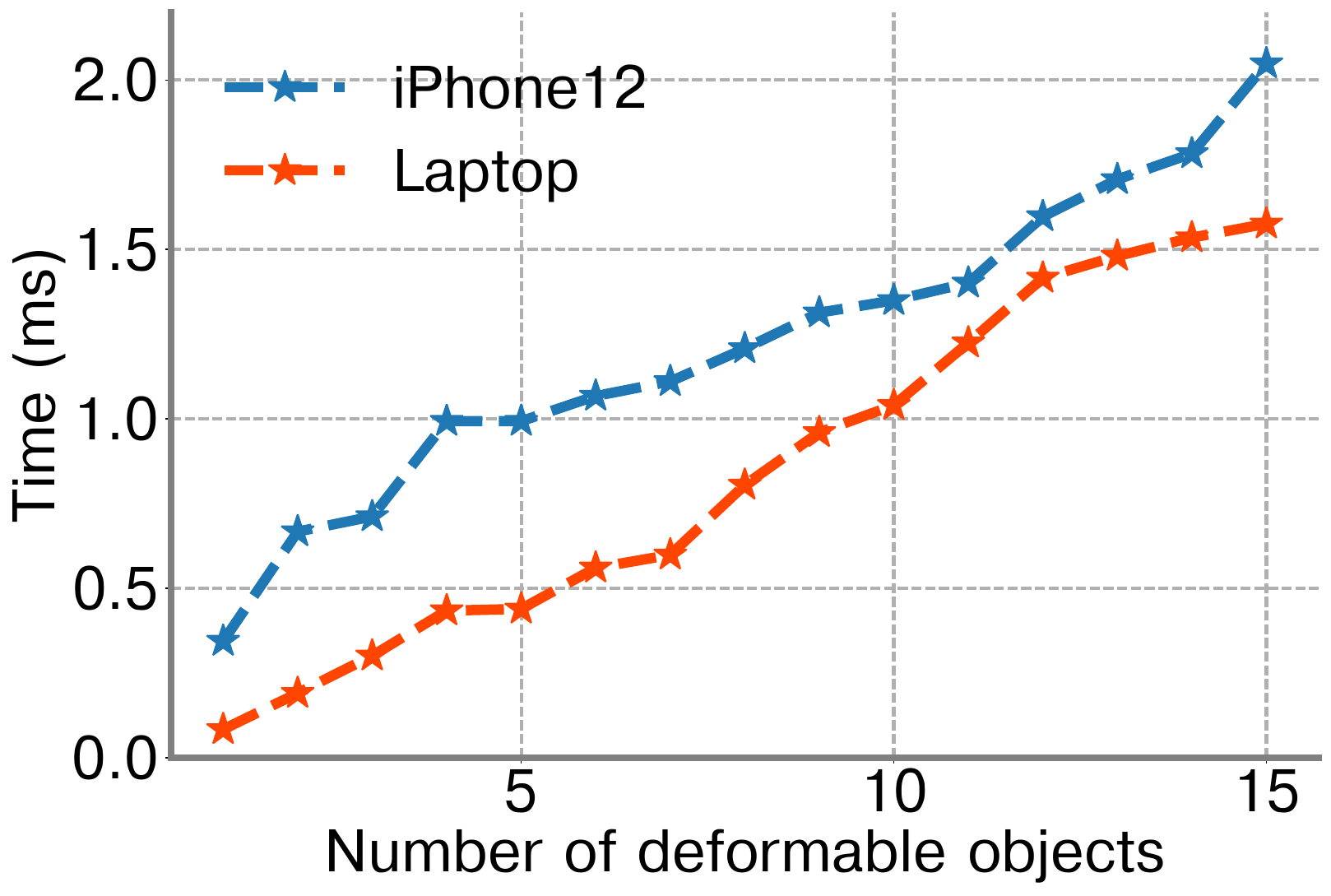}
        % \vspace{-10pt}
        \caption{Local rendering time for deformable objects.}
        % \vspace{-15pt}
        \label{fig:deformable}
    \end{figure}
\end{minipage}
\hspace{5pt}
\begin{minipage}{0.45\columnwidth}
    \begin{figure}[H]
        \centering
        % \vspace{-13pt}
        \includegraphics[width=0.75\columnwidth]{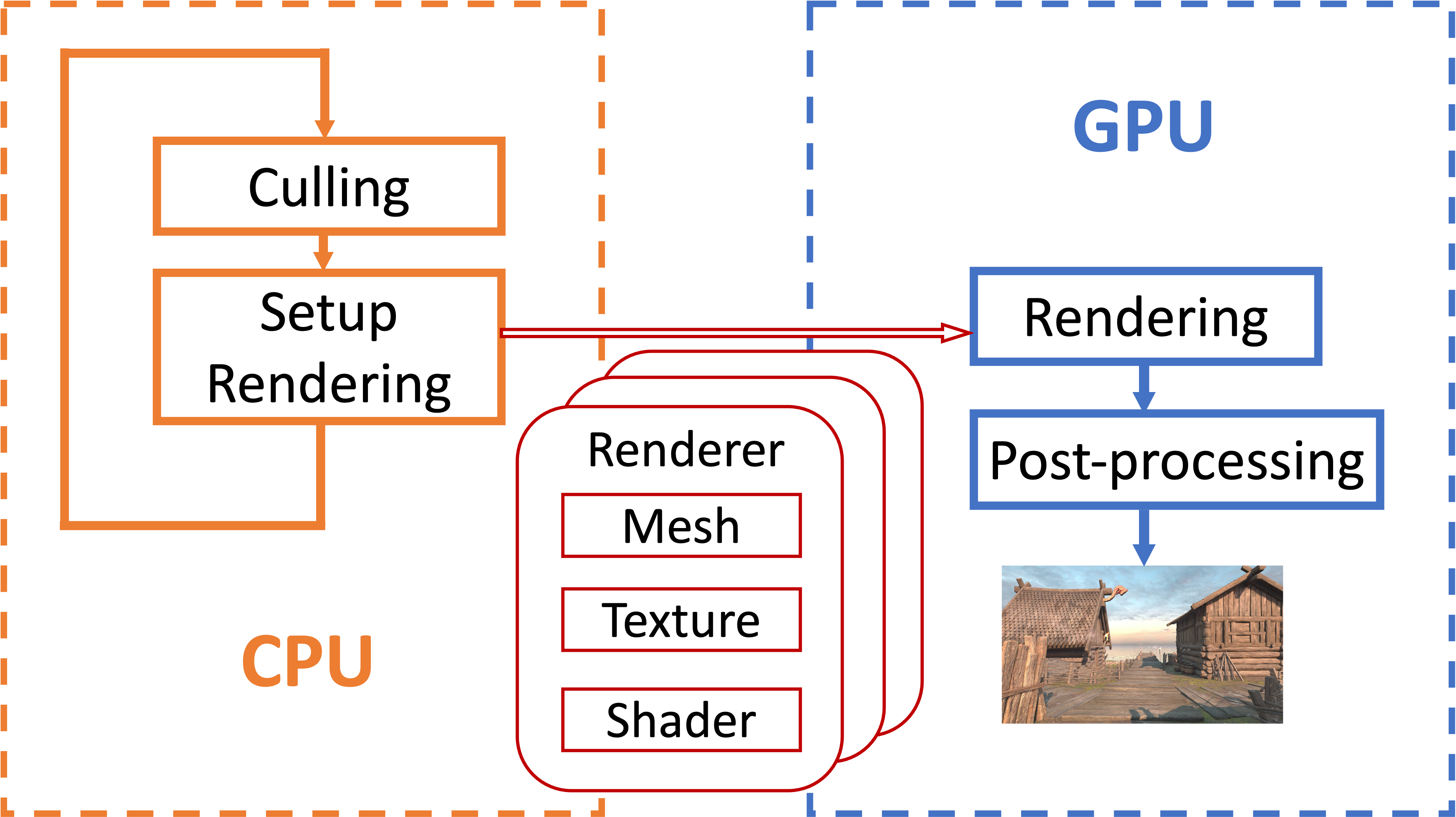}
        \vspace{14pt}
        \caption{Simplified Rendering Process in Unity}
        % \vspace{-15pt}
        \label{fig:unity_rendering_process}
    \end{figure}
\end{minipage}
\end{minipage}
\end{figure*}

% \begin{figure}[t!]
%   \centering
%   \includegraphics[width=0.8\columnwidth]{figures/network_analysis/game_duration.pdf}
%   \vspace{-10pt}
%   \caption{Duration of receiving packets for Tomb Raider, Panzer Dragoon, and Outriders.}
%   \label{fig:game_duration}
%   \vspace{-12pt}
% \end{figure}

% \begin{figure}[t!]
%   \centering
%   \includegraphics[width=0.8\columnwidth]{figures/network_analysis/game_pkt_size.pdf}
%   \vspace{-10pt}
%   \caption{CDF of video frame size for Tomb Raider, Panzer Dragoon, Outriders, and Viking Village.}
%   \label{fig:game_frame_size}
%   \vspace{-12pt}
% \end{figure}

% \vspace{-2pt}
\subsection{Foreground Interactions and Deformable Objects}
% \vspace{-2pt}

% \begin{figure}[t!]
% % \vspace{-5pt}
%     \centering
%     \includegraphics[width=0.7\columnwidth]{figures/dynamic.pdf}
%     \vspace{-10pt}
%     \caption{Local rendering time for deformable objects.}
%     \vspace{-15pt}
%     \label{fig:deformable}
% \end{figure}

\newrevised{Furion~\cite{lai2017furion} provides an insight that a game frame usually consists of foreground interactions and a background environment. The background environment is fairly predictable since it is incrementally updated as objects or camera moves in the game world. However, it has a heavier rendering load due to rich details and complex textures, so it is recommended to render and transmit from the server. The foreground interactions are more lightweight but unpredictable and hence are suited for rendering on the client side.}
 
\newrevised{We preload game objects to GPU and use our model to recover the frames that are partially or completely lost. This strategy works for rigid objects since it requires that the vertices remain the same throughout the game. If an object is deformable, we have to upload its updated vertices to the GPU again. The overhead increases with the number of deformable objects. To avoid such overhead, we also follow the cooperative rendering strategy in Furion to leverage local machine's GPU to render the deformable objects. Figure~\ref{fig:deformable} shows that the time for rendering up to 15 deformable objects is around 2ms and 1.5ms on the iPhone 12 and laptop, respectively, which is acceptable for our 30 FPS system. It suggests that we can classify the objects into rigid vs. deformable, recover the rigid objects using our model, and apply local rendering for deformable objects.}

\subsection{Summary}

In this section, we show that it is challenging to perform local rendering or video recovery in real time and the current game video prediction accuracy is insufficient. Moreover, we find that it is common to receive partially corrupted frames when packet loss happens. These observations motivate us to develop a video recovery algorithm to recover both partially corrupted frames and completely lost frames. 
% Our network trace analysis of an existing cloud gaming service shows that clients usually have idle cycles for video recovery. 
\newrevised{We follow the cooperative rendering strategy shown in Furion~\cite{lai2017furion} to leverage local machine’s GPU to render foreground interactions and deformable objects.}

\section{Extract and Represent Game States}
\label{sec:extract}

In this section, we present our approach to extracting and representing the game states. We use the games developed from Unity~\cite{unity}, but our approach is general and can potentially support other game engines. Below we first provide a brief overview of Unity and then describe our game state extraction.

\subsection{Background}

% \begin{figure}[t!]
%     \centering
%     \includegraphics[width=0.8\columnwidth]{figures/unity.pdf}
%     \vspace{-10pt}
%     \caption{Simplified Rendering Process in Unity}
%     \vspace{-15pt}
%     \label{fig:unity_rendering_process}
% \end{figure}

Unity~\cite{unity} is a cross-platform game engine developed by Unity Technologies. It is primarily used to develop video games and simulations. At a high level, the Unity rendering pipeline consists of culling, rendering, and post-processing~\cite{unity-doc}. Figure~\ref{fig:unity_rendering_process} shows a simplified rendering process in Unity. % Unity provides a few pre-built render pipelines and also allows users to create their own pipeline. 

\para{Culling:} For each frame, a camera performs frustum culling to remove renderers that fall outside the camera's view. Occlusion culling further removes the occluded objects. Both frustum culling and occlusion culling remove the need of rendering unnecessary objects and enhance efficiency. Then the CPU will upload visible renderers, camera's matrices, and others to the GPU for rendering.   

\para{Rendering:} The rendering step involves creating geometry, adding lighting effects, and projecting the 3D environment onto the 2D screen based on the camera configuration. A renderer consists of meshes, textures, and shaders. A mesh describes the geometry of a game object and is represented using vertices and multiple triangle arrays. Each game object has at least one texture, which determines its color and transparency. The shader is a program that computes the color of each pixel according to the lighting and material. 

\para{Post-processing:} Unity offers post-processing effects, such as blurring the background and moving objects, balancing colors around white areas, adding fog effects, etc. They are used to create specialized visual effects. 

\subsection{Game States}

\begin{figure*}
\centering
\begin{minipage}[ht]{1.00\columnwidth}
\begin{minipage}{0.5\columnwidth}
    \begin{figure}[H]
        \centering
        % \vspace{-10pt}
        \subfigure[Original Frame]{%
            \includegraphics[width=90px,height=50px]{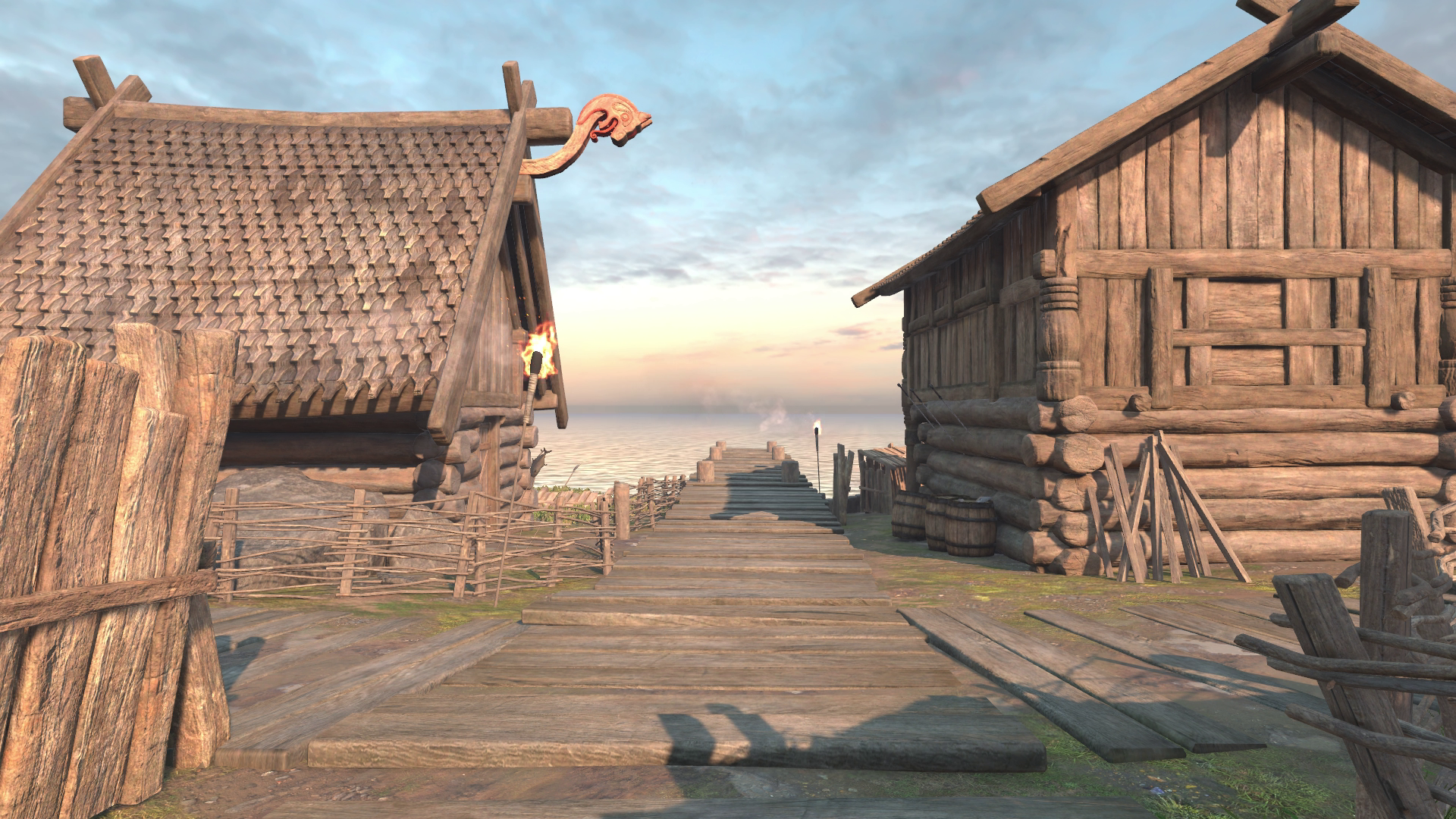}
            % \vspace{-20pt}
            \label{fig:org}
        }
        \subfigure[No Color]{%
            \includegraphics[width=90px,height=50px]{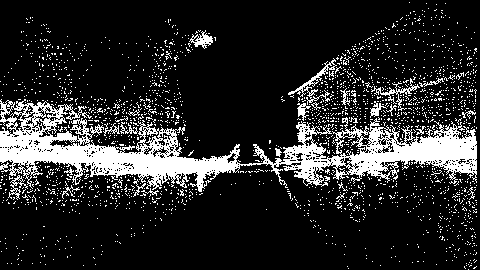}
            % \vspace{-20pt}
            \label{fig:woc}
        }
        \subfigure[Colored]{%
            \includegraphics[width=90px,height=50px]{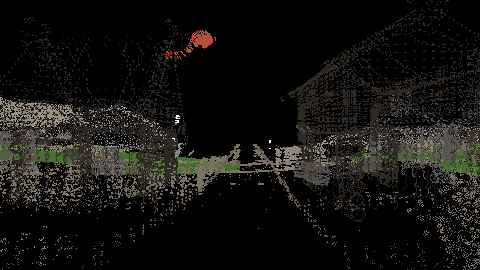}
            % \vspace{-20pt}
            \label{fig:wc}
        }
        \subfigure[Downsample 5x]{%
            \includegraphics[width=90px,height=50px]{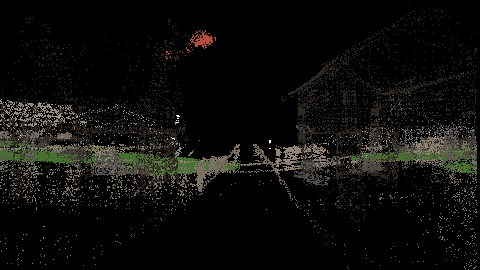}
            % \vspace{-20pt}
            \label{fig:down_5x}
        }
    % \vspace{-15pt}
    \caption{An example of game states.}
    \vspace{-10pt}
    \label{fig:example_gs}
    \end{figure}
\end{minipage}
% \hspace{1pt}
\begin{minipage}{0.5\columnwidth}
    \begin{figure}[H]
        \centering
        % \vspace{-10pt}
        \subfigure[$270\times480$]{%
            \includegraphics[width=90px,height=50px]{figures/gs_example/00010_270p.png}
            % \vspace{-20pt}
            \label{fig:270p}
        }
        \subfigure[$64\times128$]{%
            \includegraphics[width=90px,height=50px]{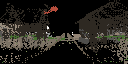}
            % \vspace{-20pt}
            \label{fig:64p}
        }
        \subfigure[$64\times64$]{%
            \includegraphics[width=90px,height=50px]{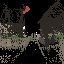}
            % \vspace{-20pt}
            \label{fig:64s}
        }
    % \vspace{-15pt}
    \caption{Game states in different resolutions.}
    \vspace{-10pt}
    \label{fig:example_gs_size}
    \end{figure}
\end{minipage}
\end{minipage}
\end{figure*}

% \begin{figure*}[htpb]
%   \centering
%   \subfigure[Original Frame]{%
%     \includegraphics[width=100px,height=60px]{figures/gs_example/00010_org.png}
%     \vspace{-20pt}
%     \label{fig:org}
%   }
%   \subfigure[No Color]{%
%     \includegraphics[width=100px,height=60px]{figures/gs_example/00010_woc.png}
%     \vspace{-20pt}
%     \label{fig:woc}
%   }
%   \subfigure[Colored]{%
%     \includegraphics[width=100px,height=60px]{figures/gs_example/00010_5x.png}
%     \vspace{-20pt}
%     \label{fig:wc}
%   }
%   \subfigure[Downsample 5x]{%
%     \includegraphics[width=100px,height=60px]{figures/gs_example/00010_270p.png}
%     \vspace{-20pt}
%     \label{fig:down_5x}
%   }
%   \vspace{-15pt}
%   \caption{An example of game states.}
%   \vspace{-10pt}
%   \label{fig:example_gs}
% \end{figure*}

We extract the game states from \newrevised{the running game app}, which include game objects' locations, speeds, rotations, sizes, etc. The game states along with the previous frames can be used together to improve the game video prediction. % We can split the Unity rendering pipeline in Figure~\ref{fig:unity_rendering_process} to efficiently generate a 2D projection from 3D mesh. Given these information and previous few frames, the video prediction model can be guided to learn a more accurate current frame. 

Algorithm~\ref{alg:game_states_extraction} shows the pseudo-code for our game state extraction. On the CPU side, we first extract all renderers and then mimic frustum culling by checking each object's bounding box against the frustum of the camera. Any renderers whose bounds are not visible will be skipped. Then we send the camera's Projection ($P$), View ($V$), and object's Transform ($T$) matrices, along with all visible vertices to the GPU. $T$ transforms from the object's local space to the global space, $V$ converts the global space to the camera view, and $P$ specifies how the camera maps a 3D world onto a 2D image. Different from the CPU processing in Figure~\ref{fig:unity_rendering_process}, we gather and send only the above information to GPU because they can completely capture the states for each frame. \newrevised{In this way, we can extract the game states without rendering the whole game frame, thereby avoiding much overhead on the client side.}
\begin{wraptable}{}{0.65\textwidth}  
    \vspace{-10pt}
    \begin{minipage}{0.65\textwidth}  
  \begin{algorithm}[H]  
    \caption{Game States Extraction}  
    \label{alg:game_states_extraction}  
    % \algsetup{linenosize=\tiny}  
    \footnotesize 
    \begin{algorithmic}[1]  
      \State \textbf{CPU:}   
        \For {$frame \,f=1,2,\ldots$}  
          \For {$object \,i=1,2,\ldots,N$}  
            \State Check the bounding box against the frustum of the camera to skip invisible objects  
          \EndFor  
          \State Send $P, V, T$ matrix and all visible vertices to the Compute Shader running on GPU  
        \EndFor  
      \newline  
      \State \textbf{GPU:}   
        \For {\textbf{all} $vertices$ in parallel}  
          \State Perform $MVP$ matrix multiplication  
          \State Transform each 3D vertex from the local space to the 2D screen space  
          \State Filter out the points outside the 2D screen  
        \EndFor  
        \State Send the game states back to CPU  
    \end{algorithmic}  
    % \vspace{-5pt}  
  \end{algorithm}  
  \end{minipage}
\end{wraptable}  
Compute shaders are programs running on the GPU outside the rendering pipeline. They can be used for massively parallel GPU computation. Since Unity does not provide APIs to access the intermediate results, we derive the game states by developing our own compute shaders to transform the 3D vertices of each visible object onto a 2D screen. To do so, we calculate the Model View Projection ($MVP$) matrix based on the camera's Projection ($P$), View ($V$), and object's Transform ($T$) matrix using $\mathbf{MVP}_{f,i} = \mathbf{P}_{f} \times \mathbf{V}_{f} \times \mathbf{T}_{f,i}$
where $f$ is the frame index and $i$ is the game object index. $MVP$ converts each object from the local space to the clip space. Then we specify the 2D screen size to perform a viewport transform and generate the final game states on the 2D screen. % Figure~\ref{fig:example_gs} shows an example, where 
Figure~\ref{fig:org} and ~\ref{fig:woc} are a RGB frame and the game state at a resolution of $270\times480$.

% no colored, colored, and downsampled game state of size $270\times480$, respectively. [XXX: are all game states 270 x 480?]

% When we start rendering an object in the game world, its vertices are used in the Vertex Processing stage to construct the shape of the object. The vertices are originally stored as 3D coordinate in the Model Space. After transformed into World Space, they can be viewed by the camera as point cloud that represents the rough shape of the object to help making the frame prediction.

As mentioned above, game state extraction involves the following steps: (i) perform culling, (ii) upload visible vertices of game objects from CPU to GPU, (iii) perform the matrix multiplication in parallel, and (iv) download the game states from GPU to CPU. % Our experiment shows that it takes XXX ms and XXX ms to generate $270\times480$ game states on the laptop and the desktop, respectively, which is even slower than local rendering. 
Equation~\ref{eq:gs_time_iphone} and ~\ref{eq:gs_time_lap} show the time for generating $270\times480$ game states in Viking Village on the iPhone 12 and laptop. 
% The profiling time on the laptop and desktop are shown in , respectively.  [XXX: explain exp]
This is even slower than performing local rendering, which is not acceptable.

% We preload vertices to significantly reduce the latency and also enrich the game states by 

\vspace{-5pt}
\begin{equation}
\label{eq:gs_time_iphone}
% {\scriptsize
\begin{aligned}
    \underset{(435.8ms)}{{T}_{gs\_iphone12}}
    = \underset{(1.5 ms)}{{T}_{culling}} + \underset{(420 ms)}{{T}_{CPU\rightarrow GPU}} + \underset{(12.8ms)}{{T}_{multiplication}} 
    + \underset{(1.5ms)}{{T}_{GPU\rightarrow CPU}}
\end{aligned}
% }
\end{equation}

\begin{equation}
\label{eq:gs_time_lap}
% {\scriptsize
\begin{aligned}
    \underset{(237.5ms)}{{T}_{gs\_laptop}}
    = \underset{(3.3ms)}{{T}_{culling}} + \underset{(230 ms)}{{T}_{CPU\rightarrow GPU}} + \underset{(4ms)}{{T}_{multiplication}} + 
    \underset{(0.2ms)}{{T}_{GPU\rightarrow CPU}}
\end{aligned}
% }
\end{equation}

\subsection{Speed-up Game State Extraction}

% To achieve 30 FPS, it is necessary to reduce each piece of the above time to meet the constraint. Meanwhile, we want to add more information to the game states for more accurate frame prediction. To this end, we make the following improvements on the game states extraction.

% \subsection{Preloading Vertices}
% \ 
% \newline 

To reduce the latency, we observe that uploading vertices from the CPU to the GPU is the most time-consuming since it involves uploading several million 3D vertices. To address this issue, we find that the vertices of game objects remain the same in the local space. Whenever the camera or game object moves, we just need to let the Computer Shader know whether the game object exists in the current frame and how the MVP matrix changes. Therefore, we preload all objects' vertices into GPU at the beginning of the game. We create an array to indicate whether a game object exists in the current frame. We finally pass the array along with the $P$, $V$, and $T$ matrices to the Compute Shader for matrix multiplication, thereby reducing ${{T}_{CPU\rightarrow GPU}}$ to $0.2ms$ and $0.08ms$ on the iPhone 12 and the laptop, respectively. The total size of all game objects' vertices is around 200MB in Viking Village and 630MB in Viking Village+, which can fit into GPU's VRAM on the iPhone 12 and the laptop.

\subsection{Game State Representation} 

\vspace{-2pt}
\para{Colored game states:} Colored game states give richer information and are likely to help video frame prediction. As an example, Figure~\ref{fig:woc} and \ref{fig:wc} show a gray-scale game state and a colored one. We can more easily see different objects from the colored game state. However, the number of I/O operations increases with the colored game state since the RGB frame requires 3 channels whereas the gray-scale has only 1 channel. For each channel, a pixel value is represented as a 4-byte floating number in Unity. Using 3 channels increases ${T}_{GPU\rightarrow CPU}$ from $1.5ms$ to $4.2ms$ on the iPhone 12. To avoid the delay increase, we build a HashMap to store the color of all game objects at the beginning of the game. In this way, the values of the game states sent back from GPU are colored indices rather than real pixels. % Each entry in the colored indices takes XXX bits. 
The CPU will look for the corresponding colors from the HashMap and rebuild the game states with RGB channels without extra cost. Note that the HashMap has a fixed number of game objects and yields a negligible increase in memory size. For example, Viking Village has around 1,500 objects and yields only 4.4KB memory increase. Moreover, we leverage the depth information of 3D vertices to filter out the farther vertex if two vertices from different game objects are mapped to the same point on the 2D screen.

% [XXX: report memory size and I/O operations]

% The colored game states is shown in Figure.~\ref{fig:5x}. 

% To assist the frame prediction process, we add colors to each game object so that our prediction model can better distinguish game objects for each frame. Taking the game states back from GPU to CPU involves I/O operations. Figure~\ref{fig:woc} is a gray scale image, which takes only one channel. Using colored game states requires us to generate RGB channels, which triples I/O operation time. To avoid this issue, we build a HashMap to store the color of all game objects before we start to play the game. In this way, the values of the game states sent back from GPU are colored indices rather than real pixels. CPU will look for the corresponding colors from the HashMap and rebuild the game states with RGB channels without extra cost. The colored game states is shown in Figure.~\ref{fig:5x}. 

% \begin{figure}
%   \centering
%   \subfigure[$270\times480$]{%
%     \includegraphics[width=70px,height=40px]{figures/gs_example/00010_270p.png}
%     \vspace{-20pt}
%     \label{fig:270p}
%   }
%   \subfigure[$64\times128$]{%
%     \includegraphics[width=70px,height=40px]{figures/gs_example/00010_64p.png}
%     \vspace{-20pt}
%     \label{fig:64p}
%   }
%   \subfigure[$64\times64$]{%
%     \includegraphics[width=70px,height=40px]{figures/gs_example/00010_64s.png}
%     \vspace{-20pt}
%     \label{fig:64s}
%   }
%   \vspace{-15pt}
%   \caption{Game states in different resolutions.}
%   \vspace{-10pt}
%   \label{fig:example_gs_size}
% \end{figure}

\para{Game state resolution:} To reduce the matrix multiplication time on GPU, we downsample the vertices in the game states. We try different downsampling ratios and measure the computation time. Table~\ref{tab:down_size_time} shows that downsampling by a factor of 2, 5, and 10 reduces ${T}_{multiplication}$ to $7.9ms$, $4.3ms$, and $3.6ms$ on the iPhone 12, and reduces to $2.5ms$, $1.3ms$, and $1.1ms$ on the laptop, respectively. 
% We choose to use a downsampling ratio of 5 since it balances the delay and accuracy well. 
Figure~\ref{fig:down_5x} shows an example of game states with $5\times$ downsampling ratio. In order to reduce ${T}_{GPU\rightarrow CPU}$, we also vary the resolution of the game states. Table~\ref{tab:down_size_time} shows that reducing the frame size from $270\times480$ to $64\times128$ and $64\times64$ reduces ${T}_{GPU\rightarrow CPU}$ to $0.8ms$ and $0.6ms$ on the iPhone 12, and reduces to $0.1ms$ and $0.08ms$ on the laptop. Figure~\ref{fig:example_gs_size} shows examples of game state frames using different resolutions. We can still clearly see the objects in all resolutions. \zhaoyuan{The game state frames are plotted in the same size just for visualization, and we use actual resolution for inference.}

% including the $64\times64$ resolution. 
% To leave enough time for predicting frames using the game states, we use the resolution of $64\times128$ for our laptop and $270\times480$ for our desktop.

\begin{table}
  \centering
  \resizebox{0.6\columnwidth}{!}{%
  \begin{tabular}{c|cc||c|cc}
    \toprule
    & \multicolumn{2}{c||}{${T}_{multiplication}$} & & \multicolumn{2}{c}{${T}_{GPU\rightarrow CPU}$} \\ \hline
    Downsampling & iPhone 12 & Laptop & Size & iPhone 12 & Laptop \\
    \midrule
    $1\times$ & 12.8ms & 4.0ms & $270\times480$ & 1.5ms & 0.2ms \\ \hline
    $2\times$ & 7.9ms & 2.5ms & $64\times128$ & 0.8ms & 0.1ms \\ \hline
    $5\times$ & 4.3ms & 1.3ms & $64\times64$ & 0.6ms & 0.08ms \\ \hline
    $10\times$ & 3.6ms & 1.1ms \\ 
    \bottomrule
    \bottomrule
  \end{tabular}
  }
  \vspace{10pt}
  \caption{${T}_{multiplication}$ with different downsampling ratios and ${T}_{GPU\rightarrow CPU}$ with different sizes}
  % \vspace{-10pt}
  \label{tab:down_size_time}
  % \vspace{-10pt}
\end{table}

\subsection{Summary}
\label{ssec:summary}

Our optimization reduces the latency of game state extraction to Equation ~\ref{eqn:gs_time_iphone2} and ~\ref{eqn:gs_time_lap2}. % Equation~\ref{eqn:gs_time_lap2} and Equation~\ref{eqn:gs_time_desk2}: 

\vspace{-5pt}
\begin{equation}
% {\scriptsize
\begin{aligned}
    \underset{(5.9-16ms)}{{T}_{gs\_iphone12}}
    = \underset{(1.5ms)}{{T}_{culling}} + \underset{(0.2ms)}{{T}_{CPU\rightarrow GPU}} + \underset{(3.6-12.8ms)}{{T}_{multiplication}} 
    + \underset{(0.6-1.5ms)}{{T}_{GPU\rightarrow CPU}}
\end{aligned}
% }
\label{eqn:gs_time_iphone2}
\end{equation}

\begin{equation}
% {\scriptsize
\begin{aligned}
    \underset{(4.5-7.6ms)}{{T}_{gs\_laptop}}
    = \underset{(3.3ms)}{{T}_{culling}} + \underset{(0.08ms)}{{T}_{CPU\rightarrow GPU}} + \underset{(1.1-4ms)}{{T}_{multiplication}}
    +   \underset{(0.08-0.2ms)}{{T}_{GPU\rightarrow CPU}}
\end{aligned}
% }
\label{eqn:gs_time_lap2}
\end{equation}

\section{Mask Generation}
\label{sec:mask_gen}

\revised{As the network performance degrades, part of video streams might be lost or corrupted. In this case, the video decoder will generate a partially corrupted frame on the client side. It is necessary to know which part of the frame is corrupted so that our recovery model can recover that part.}  

\revised{% Since the encoded video stream is not linearly correlated with the decoded frame, we cannot infer the exact corrupted part from the video stream without the knowledge of how the stream is encoded. It is well known that 
As we know, a key frame (I-frame) is a complete video frame, while a P-frame and B-frame contain the change from the reference frame. Therefore a frame can have decoding errors either because a portion of packets corresponding to the frame has corruption or because its reference frame is corrupted. In order to properly generate a mask that indicates which parts of the frame are corrupted, we need to understand the video codec.}

\revised{The widespread H.264 codec is used by both Stadia and GeForce Now, so we focus on H.264 in this paper.
% but our approach is general to any codecs.
FFmpeg, an open-source video codec, is used for our implementation. In the H.264 video stream, a coded picture consists of a number of macroblocks. They have three types. I macroblocks are predicted using intra prediction from decoded samples. P macroblocks are predicted using inter prediction from its reference frame.  B macroblocks perform similar decoding steps to P macroblocks but are predicted from one or two reference frames. All macroblocks range from $4 \times 4$ to $16 \times 16$.}

% and the prediction can be divided into $4\times4$ sub-macroblocks.  and may be divided into small partitions, up to $4\times4$ sub-macroblocks.}}  

% \textcolor{red}{fixme: In the H.264 video stream, a field (of interlaced video) or a frame (of progressive or interlaced video) is encoded to produce a coded picture. Previously coded pictures may be used for inter prediction of future frames. A coded picture consists of a number of macroblocks, each containing $16\times16$ luma samples and associated chroma samples. The macroblock has three types. I macroblocks are predicted using intra prediction from decoded samples and the prediction can be divided into each $4\times4$ sub-macroblock. P macroblocks are predicted using inter prediction from its reference picture. An inter coded macroblock may be divided into macroblock partitions (\ie, blocks of size $16\times16$, $16\times8$, $8\times16$, or $8\times8$). If the $8\times8$ partition size is chosen, each $8\times8$ sub-macroblock may be further divided into sub-macroblock partitions of size $8\times8$, $8\times4$, $4\times8$, or $4\times4$. B macroblocks perform similar decoding steps to P macroblocks but are predicted from one or two reference pictures.}}

\begin{figure}
  \centering
  \subfigure[Corrupted Frame]{%
    \includegraphics[width=0.26\columnwidth]{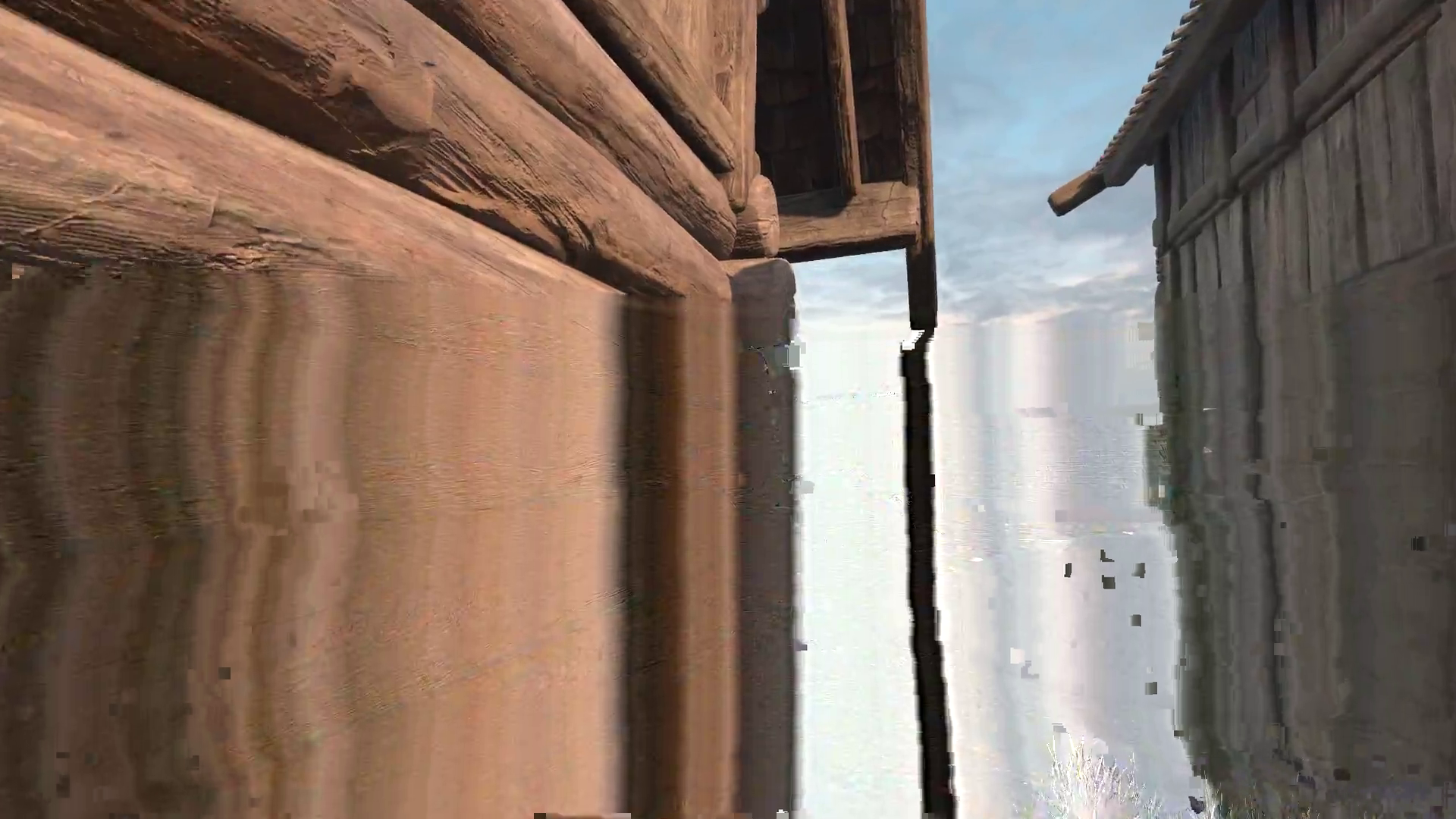}
    % \vspace{-20pt}
    \label{fig:partial}
  }
  \subfigure[Mask]{%
    \includegraphics[width=0.26\columnwidth]{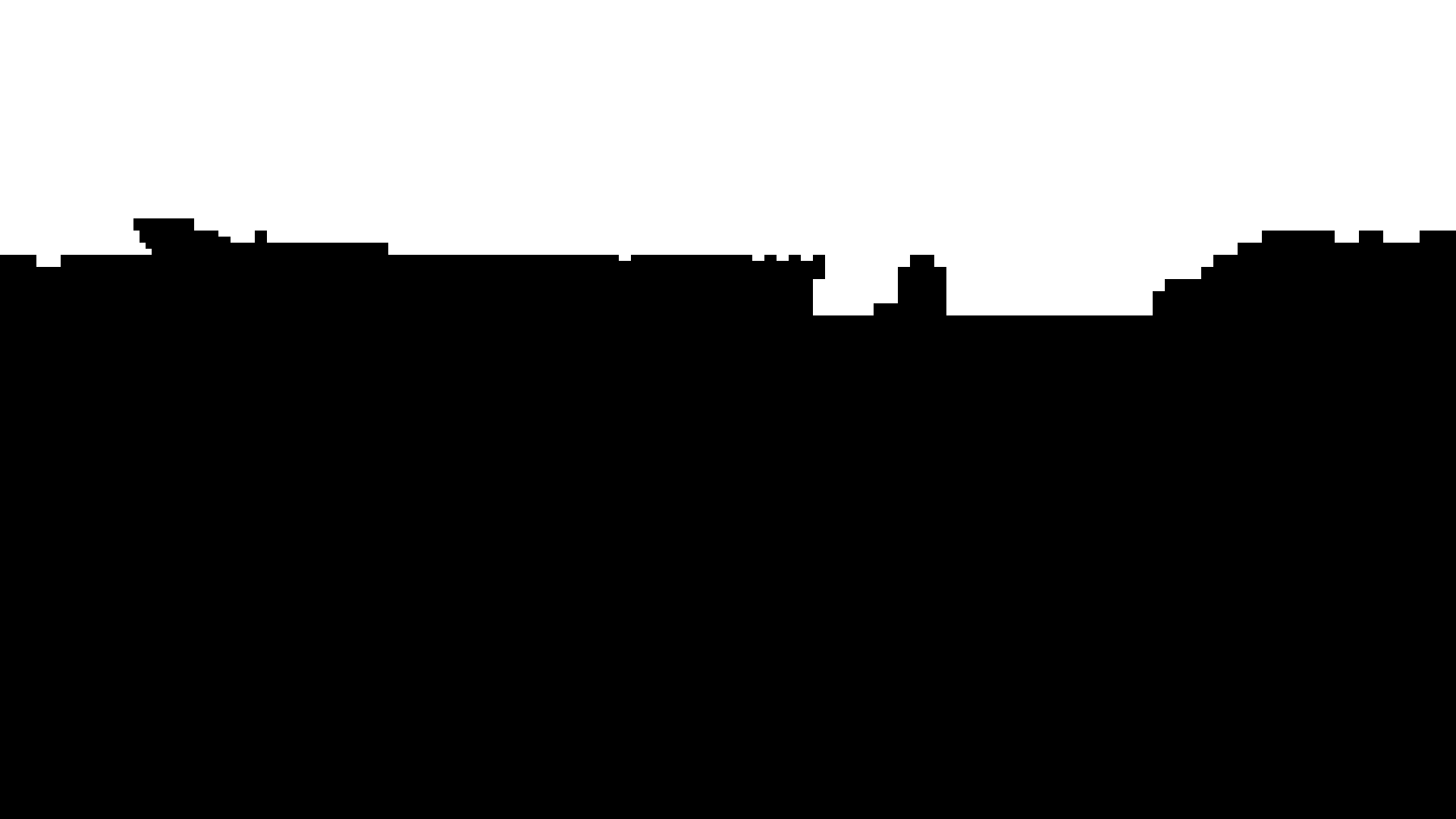}
    \vspace{-20pt}
    \label{fig:mask}
  }
  % \vspace{-15pt}
  \caption{An example of a corrupted frame and the corresponding mask.}
  % \vspace{-15pt}
  \label{fig:example_mask}
\end{figure}

\revised{The H.264 decoder will not continue to decode the subsequent macroblocks if previous ones are missing or corrupted, thereby giving us an accurate position indicating which macroblocks are corrupted. The minimum size of sub-macroblocks is $4\times4$ and we can always get their reference blocks from inter or intra prediction during decoding, so we generate a mask to indicate whether each sub-macroblock can be decoded correctly. When generating the mask, we take into account both the macroblock itself and its reference block. 
% accurate even if a P frame is fully received but its reference frame is corrupted. 
Figure~\ref{fig:example_mask} shows an example of a corrupted frame and its mask. In the mask, the white color denotes a correctly decoded part while the black color denotes a corrupted part.}

\revised{With the mask, we can leverage partially decoded frames for video recovery. It incurs a negligible overhead because we only use intermediate variables and do not require additional computation during the decoding. The mask is a gray-scale array that takes around 2MB of memory. The delta frame might be a few frames away from its reference frame, so we need to cache several masks (\eg, 10) in the memory for future reference.}

% \section{Leverage Game States}
\section{Video Recovery Model}
\label{sec:leverage}

\begin{figure}[htbp]
  \centering
  % \vspace{-10pt}
  \includegraphics[width=0.85\columnwidth]{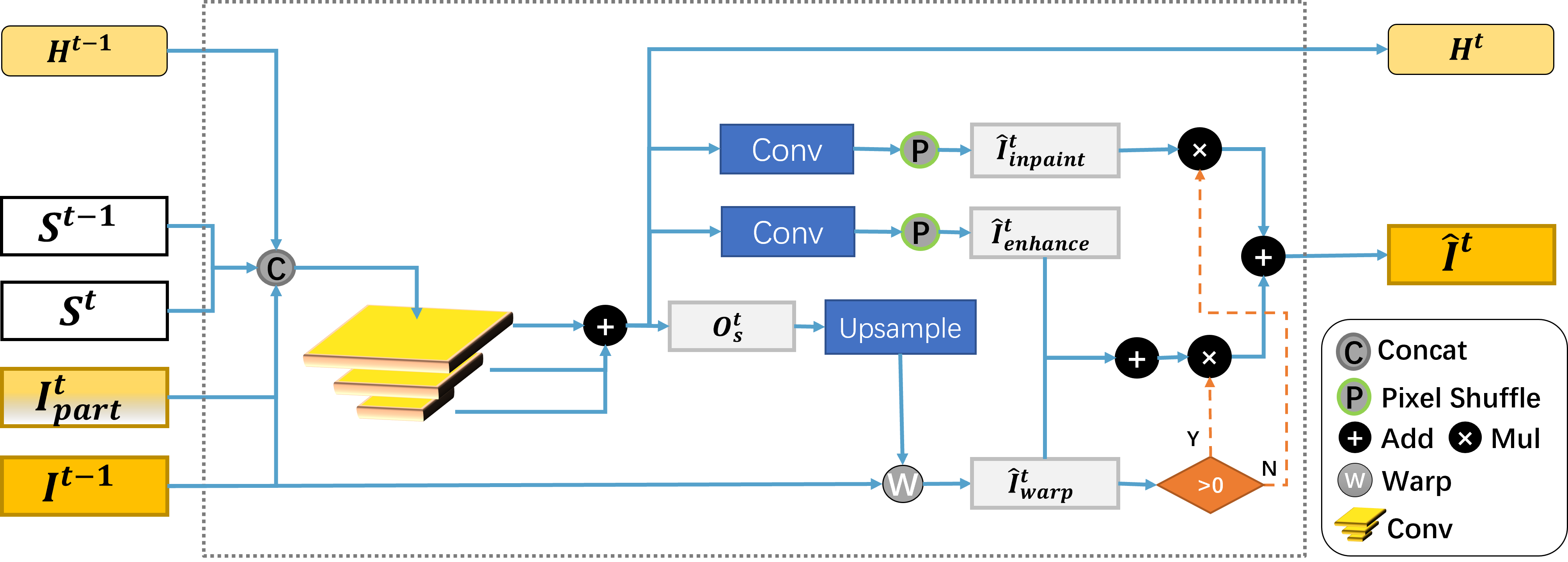}
  % \vspace{-10pt}
  \caption{Model architecture diagram. }
  % \vspace{-10pt}
  \label{fig:arch}
\end{figure}

A natural way to recover a video frame is to use the previous frames. There have been lots of existing works in this direction (\eg, \cite{Yu2019CrevNetCR, Kaur2020FutureFP,video-pred-survey}). In cloud gaming, we can further leverage the current game states in addition to the previous video frames to improve the performance since the game states contain important information regarding the current objects' shapes, sizes, colors, orientations, positions, movements, etc. 

% The way to get game state information is simple but efficient. It contains lots of information that is impossible to get with traditional future frame prediction methods~\cite{Yu2019CrevNetCR, Kaur2020FutureFP}. The game state information directly describes the exact distribution of objects location and color for the next frame, making prediction considerably easier.

\yifan{Game state changes strongly correlate with RGB frames (from Unity's rendering logic), so they help our network to learn the mapping to RGB space optical flow.} Our intuition is that the pixel-wise shifting $\mathbf{O}_{S}^t$ between the two game states $\mathbf{S}^{t-1}$ and $\mathbf{S}^t$ at time $\mathit{t}$ is similar to $\mathbf{O}_{I}^t$ between the corresponding two frames $\mathbf{I}^{t-1}$ and $\mathbf{I}^t$. Therefore, we can first compute $O_S^t$ between the current and previous game states, and use it to warp the previous frame $I^{t-1}$ to predict the current one $I^t$. That is, we compute
$\mathbf{O}_{s}^t = \mathcal{O}(\mathbf{S}^{t-1}, \mathbf{S}^t)$ and $\mathbf{\widehat{I}}^t = \mathcal{W}(\mathbf{I}^{t-1}, \mathbf{O}_{s}^t)$, where $O()$ denotes the optical flow and $W$ denotes warping.

%More formally, we have: 

% If we have $\mathbf{S}^t$, we can easily get $\mathbf{O}_{s}^t$ and use it to warp $\mathbf{I}^{t-1}$ to get the predicted $\mathbf{\widehat{I}}^t$ of the next frame. Therefore, we have the following:

%\begin{equation}
%    \mathbf{O}_{s}^t = \mathcal{O}(\mathbf{S}^{t-1}, \mathbf{S}^t)
%\label{eqn:ost}
%\end{equation}
%\begin{equation}
%    \mathbf{\widehat{I}}^t = \mathcal{W}(\mathbf{I}^{t-1}, \mathbf{O}_{s}^t)
%\label{eqn:it}
%\end{equation}

% \begin{figure*}[htpb]
%   \centering
%   \subfigure[Last Frame]{%
%     \includegraphics[width=90px,height=50px]{figures/optical_flow/last.png}
%     \vspace{-20pt}
%     \label{fig:last}
%   }
%   \subfigure[Current Frame]{%
%     \includegraphics[width=90px,height=50px]{figures/optical_flow/gt.png}
%     \vspace{-20pt}
%     \label{fig:current}
%   }
%   \subfigure[Optical Flow]{%
%     \includegraphics[width=90px,height=50px]{figures/optical_flow/warp_flow.png}
%     \vspace{-20pt}
%     \label{fig:of}
%   }
%   \subfigure[Refined Flow]{%
%     \includegraphics[width=90px,height=50px]{figures/optical_flow/warp_refined_flow.png}
%     \vspace{-20pt}
%     \label{fig:rf}
%   }
%   \subfigure[Generation]{%
%     \includegraphics[width=90px,height=50px]{figures/optical_flow/pred_3.png}
%     \vspace{-20pt}
%     \label{fig:rg}
%   }
%   \vspace{-10pt}
%   \caption{Optical Flow vs Refined Flow.}
%   \label{fig:opticalflow}
%   \vspace{-10pt}
% \end{figure*}

\subsection{Estimating Optical Flow}

The optical flow captures the movement of the brightness pattern and approximates the motion field. Let $u(x,y)$ and $v(x,y)$ denote the 2D optical flow for a given point $(x,y)$ in a video frame.  Under the assumption of the "intensity constancy", we have 
$I(x+\Delta x, y+\Delta y, t+\Delta t) \approx I(x,y,t).$
Through a Taylor series expansion, 
$I(x+\Delta x, y+\Delta y, t+\Delta t) = I(x,y,t) + \frac{\delta I}{\delta x} \Delta x + \frac{\delta I}{\delta y} \Delta y + \frac{\delta I}{\delta t} \Delta t.$ Hence we have 
$ \Delta I v + I_t = 0$
where $v = (\Delta x, \Delta y)$ is the optical flow, $\Delta I = (I_x, I_y)$ is the spatial gradient, and $I_t$ is the temporal gradient. By deriving $\Delta I$ and $I_t$ from a pair of video frames, we get the optical flow $v$. 

Optical flow can be estimated by phase correlation, block-based, differential methods, discrete optimization, and deep learning approaches. We choose SPyNet~\cite{ranjan2017optical}, one of the most widely used optical flow networks due to its high efficiency and accuracy. 

% [XXX: describe the network, downsamples by a factor of 32] 
% due to its low cost, which is necessary to support our real-time requirements. [XXX: reason to use this particular optical network]

Figure~\ref{fig:arch} shows our model architecture. We initialize the pyramidal convolution backbone with a SPyNet pre-trained on Sintel~\cite{Butler:ECCV:2012}. It uses the current game states $\mathbf{S}^t$ and the previous one $\mathbf{S}^{t-1}$ to capture optical flow. It then downsamples the input $5$ times through a pyramid structure and accumulates the predicted optical flow layer by layer to get $\mathbf{O}_{S}^t$.

\subsection{Refining \& Warping Using Optical Flow}

% The optical flow between the game states can be used to warp the previous RGB frame to the current RGB frame. Essentially for each pixel $p$ in the previous RGB frame, we move it to a new position $p + \mathbf{O}_{s}^t(p)$.  

% The optical flow output $\mathcal{O}_W$ can be used to warp the previous RGB frame to the next RGB frame according to Equation~\ref{eqn:it}. Warping an image is essentially performing a 2D interpolation. [XXX：forward or backward warping]

% [XXX: warp formula]

% Note that the size of game states and RGB frames are not necessarily the same size, so it is necessary to use Bilinear Interpolation to up-sample the resulting optical flow while the pixel offset values are scaled equallyi.

% \subsection{Refined Flow for the Information Gap}

The optical flow between the two consecutive game states can be used to warp the previous RGB frame to the current one. However, the optical flow between the two game states differs from the optical flow between the two corresponding RGB frames in the following way: (i) game states have a finite number of scatter points and their frames are more sparse, (ii) the movement of the scatter points are not the same as the movement of brightness patterns in RGB frames, \revised{and (iii) the resolution of game states is usually much lower than RGB frames for the sake of computational efficiency, so the resulting flow cannot be used directly for warping.}

\yifan{For the output optical flow $\mathbf{O}_{S}^t$, we use bilinear interpolation and grid sample-based warping to create an aligned prediction $\mathbf{\widehat{I}}_{warp}^{t}$ from the previous frame $\mathbf{I}^{t-1}$. This fully differentiable process enables gradient propagation to  optimize the optical flow network and produce accurate results.}
% Therefore, directly applying the optical flow between the game states for warping the RGB frame can lead to significant distortion. 

% as $\mathcal{O}_{W^'}$ in Equation~\ref{eqn:it} to generate a special optical flow, whose goal is not to generate the optical flow between the two game states, but to learn to closely approximate the optical flow between the RGB frames.  Specifically, we  use the following objective function: 

% make the pixels of $\mathbf{S}^{t-1}$ as close as possible to the pixel offset information of $\mathbf{S}^{t}$. [XXX: not sure about the prev sentence]

% [XXX: double check]
% The size of our game states and image frames are not equal, the generated optical flow needs to be up-sampled to transform the RGB frames. We use Bilinear Interpolation to up-sample the resulting optical flow while the pixel offset values are scaled equally.
% [XXX: add a formula]

% It can warp the previous frame $\mathbf{I}^{t-1}$ directly to get the predicted frame as $\mathbf{\widehat{I}}^t$. 

\vspace{-2pt}
\subsection{Enhancement and Inpainting\label{sec:Generate}}

Warp-based video frame prediction can only predict the visible content in the previous RGB frames, and cannot modify the color gap for the new frame or generate content that does not exist in previous frames. Therefore, we improve the optical flow network by adding the following inputs to enable the generation of new content: (i) previous frame $\mathbf{I}^{t-1}$ to add unavailable details in the game states, (ii) current partially decoded frame $\mathbf{I}_{part}^{t}$ as a reference to align the generated content to the received partial frame, and (iii) hidden state $\mathbf{H}^{t-1}$, including historical information to make the generated frames temporally smooth.
 
\revised{After the pyramid network, we generate $\mathbf{\widehat{I}}_{enhance}^{t}$ and $\mathbf{\widehat{I}}_{inpaint}^{t}$ using two sets of convolutions as Figure \ref{fig:arch} shows. We use pixel shuffle\cite{shi2016real} to scale the feature map to the size of RGB frames. $\mathbf{\widehat{I}}_{enhance}^{t}$ learns the gap from $\mathbf{\widehat{I}}_{warp}^{t}$ to the ground truth to make the prediction more natural. It focuses on the meaningful content of the warped image and ignores the empty parts. $\mathbf{\widehat{I}}_{inpaint}^{t}$ generates new content when warping cannot find reference in previous frames, so it is only responsible for the empty part of $\mathbf{\widehat{I}}_{warp}^{t}$. The final prediction can be expressed as follows: \[\mathbf{\widehat{I}}^t = (\mathbf{\widehat{I}}_{warp}^{t} + \mathbf{\widehat{I}}_{enhance}^{t}) * {M} + \mathbf{\widehat{I}}_{inpaint}^{t} * (1 - {M})\]
where ${M}$ is a bitmap and its value is 1 if $\mathbf{\widehat{I}}_{warp}^{t}$ is not empty otherwise 0. Finally, we overwrite it with $\mathbf{I}_{part}^{t}$ that has already been decoded.}
 
We quantify the difference between the predicted and actual RGB frames using the following loss function: 
% so that its goal is not to generate the optical flow between the %two game states, but to approximate the optical flow between the corresponding RGB frames. We  use the following objective function: 
%\begin{equation}
$\mathcal{L}^t = \mathcal{L}_{Pix}^t + \alpha\mathcal{L}_{distill}^t$. $\alpha$ is set to 0.1. $\mathcal{L}_{Pix}^t = \mathcal{C}(\mathbf{\widehat{I}}^t, \mathbf{I}^t)$
%\label{eqn:lpix}
%\end{equation}
%\begin{equation}
 and   $\mathcal{C}(\mathbf{\widehat{I}}^t, \mathbf{I}^t) = \sum_{\mathbf{i}}^{\mathbf{W}}\sum_{\mathbf{j}}^{\mathbf{H}}\sqrt{(\mathbf{\widehat{I}}^t_{ij} - \mathbf{I}^t_{ij})^{2} + \epsilon^{2}}$
%\label{eqn:lpix}
%\end{equation}
where $\mathbf{\widehat{I}}^t$ and $\mathbf{I}^t$ denote the actual and predicted frames at time $t$, respectively. $\mathbf{W}$ and $\mathbf{H}$ denote the width and length of the input video, and $\epsilon$ is a small constant set to $1e^{-12}$. We use Charbonnier loss~\cite{Lai2019FastAA}  as the loss function $\mathcal{C}$.
% to get $\mathbf{\widehat{I}}^t$ and $\mathbf{I}^t$ close. 
The Charbonnier loss is a differentiable variant of $\mathcal{L}_1$ distance, which is shown to perform well for generation tasks~\cite{Barron_2019_CVPR,9448108,dai2020awnet}. \revised{% Since we add an additional output term to the optical flow network, 
We also add a distillation loss\cite{shi2016real,kong2022ifrnet} $\mathcal{L}_{distill}^t = \mathcal{C}(\mathbf{O}_{S}^t, \mathbf{O}_{fix}^t)$ to the output optical flow $\mathbf{O}_{S}^t$. $\mathbf{O}_{fix}^t$ comes from the output of the fixed SPynet we used for initialization in order to maintain the warping stability. We find that without this item, $\mathbf{O}_{S}^t$ tends to spread the previous frame over the entire image to reduce loss, which renders the inpainting module ineffective. % Putting together, the overall loss is $\mathcal{L}^t = \mathcal{L}_{Pix}^t + \alpha\mathcal{L}_{distill}^t$, where $\alpha$ is set to 0.1.}

\section{System Implementation}
\label{sec:system}

\revised{Figure~\ref{fig:system_arch} shows our system architecture. The game client has a running game app to efficiently generate game states without rendering the whole game frame when users play the game. The video decoder is utilized to generate a mask to indicate which portions of video frames need recovery.} The game states and corrupted frames are fed to our recovery model for video frame recovery. Our goal is to develop a system that can be easily applied to different games. Meanwhile, in order to satisfy the 30 FPS requirement, we need to keep the total processing time within $33ms$. We also devise a strategy about when to run the recovery model and how to optimize video quality.

\zhaoyuan{\para{Easy to deploy:} Our game states extraction is easy to deploy. We provide two scripts. The first one automatically identifies and preloads vertices into the GPU during initialization and transmits matrices for extracting game states for each frame. The second script uses a Compute Shader on the GPU to generate the game states. These scripts can be used for any new games without modification. Programmers simply need to attach the scripts as modules to the game’s main camera through the Unity UI. This can be done by double-clicking on the main camera and importing the script without understanding the game’s source code. Furthermore, we modify video codec to generate masks, which can be compiled into FFmpeg during video decoding.}

\para{Speeding up inference time:} \revised{The inference time for our recovery model highly depends on the game state size. We choose the size of $64\times128$ because it gives us a good balance between delay and accuracy as shown in Section~\ref{sec:eval}. % The impact of the size of game states will be evaluated in the next section. 
Our target devices are iPhone 12 and laptop, which have different computational capabilities and we use different speedup methods on these devices.}

% Our neural recovery model adapts to varying device computation costs by using feature map sizes independent of output sizes, and applying an optical flow up-sampling mechanism to enhance RGB images at any resolution. The computational load relies on the optical flow feature map size. Note that we do not need to generate feature maps as large as output resolution. Instead, feature maps adapt according to device performance to optimize usage without limiting output sizes.

\yifan{Our neural recovery model is highly flexible and can support various feature map sizes as input to accommodate different hardware constraints. In order to generate the target output size, we apply an optical flow upsampling to the feature map with different upsampling rates to match the target output size. This allows us to freely choose the size of the feature map for generating optical flow, thus making efficient use of the device's computational power.}

%\para{iPhone12: } 
\revised{We use CoreML as the model format on the iPhone 12 because it optimizes on-device performance for iOS by leveraging the CPU, GPU, and Neural Engine. Therefore, we convert pretrained recovery model from Pytorch to CoreML and it performs faster than ONNX, Pytorch Mobile, and TensorFlow Lite. However, we observe that the grid sample operation of warping runs slowly because it is not officially supported by GPU. To address the above issue, we leverage Metal Performance Shaders (MPS) to create a custom grid sample layer running on GPU. MPS is a framework with handy Metal compute kernels and CoreML uses it for model inference on GPU. It also provides us with many APIs to create a custom layer so that the grid sample is implemented with a GPU acceleration on the iPhone 12.
In addition, we perform warping at a smaller scale of 270p instead of 1080p, thereby reducing the warping time from $29ms$ to $5ms$. The lost texture details caused by the downsampling will be made up by the enhancement module in Section~\ref{sec:Generate}. To further reduce the inference time, we use FP16 precision for both inputs and model weights without performance degradation. The final inference time is $22ms$ on the iPhone 12.}

%\para{Laptop: } 
\revised{Since our laptop runs Linux and has Nvidia GPU, we can use Pytorch with CUDA as the inference framework. To optimize the inference speed, we choose TVM~\cite{chen2018tvm}, which uses compiler techniques to optimize model inference. TVM works well with Pytorch and provides us with a 50\% inference time reduction. The final inference time on the laptop is $25ms$.}

\para{Scheduling recovery: } A natural solution is to run video recovery upon a partial or complete loss of a video frame. However, this yields large delay since the client's game state extraction and recovery can only happen after the current video frame times out. To reduce the delay, we prefer overlapping the client side game state extraction with the server side processing and network transmission. But this requires the client extract the game state without knowing whether the current video frame will arrive in time. To limit the client's resource consumption and end-to-end delay, we let the client start game state extraction for a video frame if the previous video frame is lost since network losses are bursty according to both previous measurement studies~\cite{xu2020understanding, yu2014can} and our own measurement. By analyzing the network traces shown in Table~\ref{tab:network_traces}, we calculate the loss probability of the current frame conditioned on the previous frame being lost is around 95\%.

% Our recovery model is designed to handle situations where a video frame is either partially or completely corrupted. In such cases, the client's game state extraction may coincide with network delays and video encoding/decoding time, while the DNN inference introduces an additional delay. While the most straightforward approach would be to reconstruct every video frame, regardless of whether it suffered any loss, this would result in unnecessary computational resource utilization. To tackle this challenge, we adopt a different strategy: we reconstruct the current video frame only when the preceding frame is lost or corrupted. Conversely, if the previous frame is successfully received, we skip the model inference process for the current frame. To justify this strategy, we calculate the probability of the current frame being corrupted or lost, given that the previous frame experienced loss or corruption. This probability exceeds 95\%, due to the common bursty distribution of network loss and the high inter-frame dependencies imposed by the video codec. Leveraging our neural recovery approach, if a part or complete video frame can be decoded before the deadline, we replace the corresponding portion of the predicted frame with the decoded version.
}

% Our recovery model should only be called when a frame is partially or completely corrupted, but there is a possibility that the duration of receiving packets is larger than 7ms and 5ms on the iPhone12 and laptop. In this case, the running time will exceed the budget if we call the model after detecting a corrupted frame. To address this issue, we set a timeout to 7ms and 5ms on the iPhone12 and laptop. Upon timeout, we execute our video recovery. If a part or complete video frame can be decoded before the deadline, we will replace the portion of the predicted frame with the decoded version. Using a timeout may unnecessarily execute our video recovery for the frames that can be decoded. From our experience, this is rare (\eg, only 1\% frames are received but cannot be decoded within the timeout on the iPhone 12 and laptop). 

% 7 ms on the iPhone12 or within 5 ms on laptop). 

% make sure that the model will be always called after the timeout. If the video decoding cannot finish before that, our recovery model will predict a whole frame. Besides, if part of the frame eventually can be precisely decoded, we will overwrite the predicted frame with the decoded part under the guidance of the mask to maximize the video quality.

\zhaoyuan{
\para{End-to-end latency: }With our system strategies, we can perform neural recovery without waiting for video packets received, so the recovery latency would be the sum of game states extraction runtime and model inference time. According to Section~\ref{ssec:summary}, we have a range of runtime for game states generation. On the iPhone 12, we choose the game state resolution of $64\times128$ and downsample it by $5\times$, resulting in $7ms$ for game states generation and a total latency of $29ms$. On the laptop, we do not downsample the game states and utilize a resolution of $270\times480$, which leads to $5ms$ for game states extraction and a total latency of $30ms$.
}

\eurosys{
The end-to-end latency in cloud gaming refers to the duration starting from when the user enters an input to when the resulting frame is displayed on the client device. Users can tolerate an end-to-end delay ranging from $80-150ms$~\cite{geforce_now_sys_req, choy2014hybrid, wu2016streaming, chen2019framework} depending on game types, so we choose $80ms$ as the maximum latency budget. To ensure the latency with recovery less than the time budget, we set a timeout before executing video recovery. If the partially corrupted frame can be received and processed within the timeout, our system will perform video concealment to recover the frame. Otherwise, the video prediction will be executed to avoid passing the deadline. 
}

\section{Performance Evaluation}
\label{sec:eval}

In this section, we first present the evaluation methodology and then describe the performance results.

\subsection{Evaluation Methodology}
\label{ssec:eval-method}

\vspace{-2pt}
\para{Datasets and metrics: } We use Viking Village~\cite{vk}, Nature~\cite{nature}, Corridor~\cite{corridor} on the iPhone 12, and Viking Village+ on the laptop. Each game video lasts around 5 mins and they are collected by different players. There are 50,000 training frames and 10,000 testing frames for each game. We quantify the quality of our recovered video frames using two widely used video quality metrics: SSIM and PSNR. 
% To evaluate generation, we use two additional measures widely used for generation tasks: FID\footnote{https://github.com/mseitzer/pytorch-fid}\cite{Heusel2017GANsTB} and LPIPS\footnote{https://github.com/richzhang/PerceptualSimilarity}\cite{zhang2018perceptual}. 
Higher SSIM and PSNR values indicate better video quality. 

\begin{table}
  \centering
  \resizebox{0.48\columnwidth}{!}{%
  \begin{tabular}{c|c|c|c|c}
    \toprule
    & 4G & 5G & WiFi & LEO \\
    \midrule
    Amount of traces & 145 & 152 & 150 & 146 \\
    Avg. Duration (s) & 320 & 390 & 300 & 300 \\
    Avg. Throughput (Mbps) & 32.5 & 61.7 & 72.8 & 28.9 \\
    Avg. Loss rate (\%) & 3.2 & 2.3 & 0.9 & 9.4 \\
    \bottomrule
  \end{tabular}
  }
  \vspace{10pt}
  \caption{Network traces}
  % \vspace{-12pt}
  \label{tab:network_traces}
  % \vspace{-10pt}
\end{table}

\para{Network traces: } \newrevised{Table~\ref{tab:network_traces} shows the network traces we use. We measured the downlink throughput using {\em iperf} from an Azure server located in the central U.S. to a local client over the Internet, where we varied the wireless hop to use WiFi, 4G, and 5G networks. {\em iperf} generates TCP data streams and provides the number of retransmissions and throughput. We compute the packet loss rate by dividing the number of retransmissions by the total number of transmissions. The 4G and 5G traces include static and walking scenarios. We also move the local client randomly to add mobility to the WiFi traces. The low earth orbit (LEO) satellite network traces were collected from the StarLink network in January 2023 using a StarLink RV ground station in the west coast of U.S. under a static scenario. To measure the packet loss rate, we ping the server every $10ms$ and collect the average ping drop rate.}

% We capture a large number of network packets while playing three cloud games: Tomb
% Raider, Panzer Dragoon, and Outriders on Stadia. Each game lasts for around 600 seconds. We compute the average throughput and use it as the fixed sending rate for our experiments. Then we collect packet-level network traces by encoding 30 FPS game videos using H.264 and sending the resulting videos at the fixed rate from an Azure server to a local client over the Internet. Next, we add cross-traffic to the network to evaluate the performance under different network conditions. The traces contain the timestamp for each received packet.

% \para{Baseline comparison:} \textcolor{red}{Yifan please revise this paragraph. Maybe this paragraph can be deleted and the schemes can be introduced in the Impact of game states? } We compare the following schemes for our laptop: (i) reusing the previous frame $I^{t-1}$, (ii) Recovery through $I^{t-2}$ and $I^{t-2}$, (iii) using the optical flow between $I^{t-1}$ and $I^{t-2}$ to warp $I^{t-1}$ (\ie, downsampling $I^{t-1}$ and $I^{t-2}$ to match the game state frame size and use them as the input), (iv) using the optical flow between $S^{t-1}$ and $S^t$ to warp $I^{t-1}$, and (v) using the refined flow between $S^{t-1}$ and $S^t$ to warp $I^{t-1}$ (our final scheme for our laptop). For our \textcolor{red}{desktop}, we compare the schemes with and without game states as well as different loss functions in the generation network. These comparisons shed light on the benefits of using game states and different ways of using it.

\para{Evaluation strategies: }\revised{We use the collected network traces to delay packets according to the timestamp recorded by the traces before sending them to the client. The client executes our system by decoding the video and recovering corrupted frames if needed. We compare the video recovery in two ways. First, we recover every frame assuming all previous frames are correctly received. Second, we recover the corrupted frame with previously rendered or recovered frames. In this case, the recovered frame will be fed to the model for the next frame recovery. These two strategies are complementary: the former focuses on the recovery for different frames while the latter focuses on the impact of realistic network losses. Note that we use network traces for evaluation so that all recovery algorithms are evaluated under the same loss conditions. We consider a lost packet when it misses its deadline where the deadline is determined based on the maximum tolerable end-to-end latency, i.e. $80ms$, plus the inter-frame spacing (\eg, 33 ms for 30 FPS).}
\para{Implementation details: } % We collect XXX cloud gaming traces for each of the games. We partition them into 50,000 training sets and 10,000 test sets. 
We use 100,000 iterations for all training rounds. We use an 8 GPU Tesla V100 machine with a batch size of 16. We set the learning rate to $1e^{-5}$, and use the cosine decay. 

\subsection{Performance Results}
\vspace{-3pt}

\begin{figure}[t!]
    \centering
    \includegraphics[width=0.9\columnwidth]{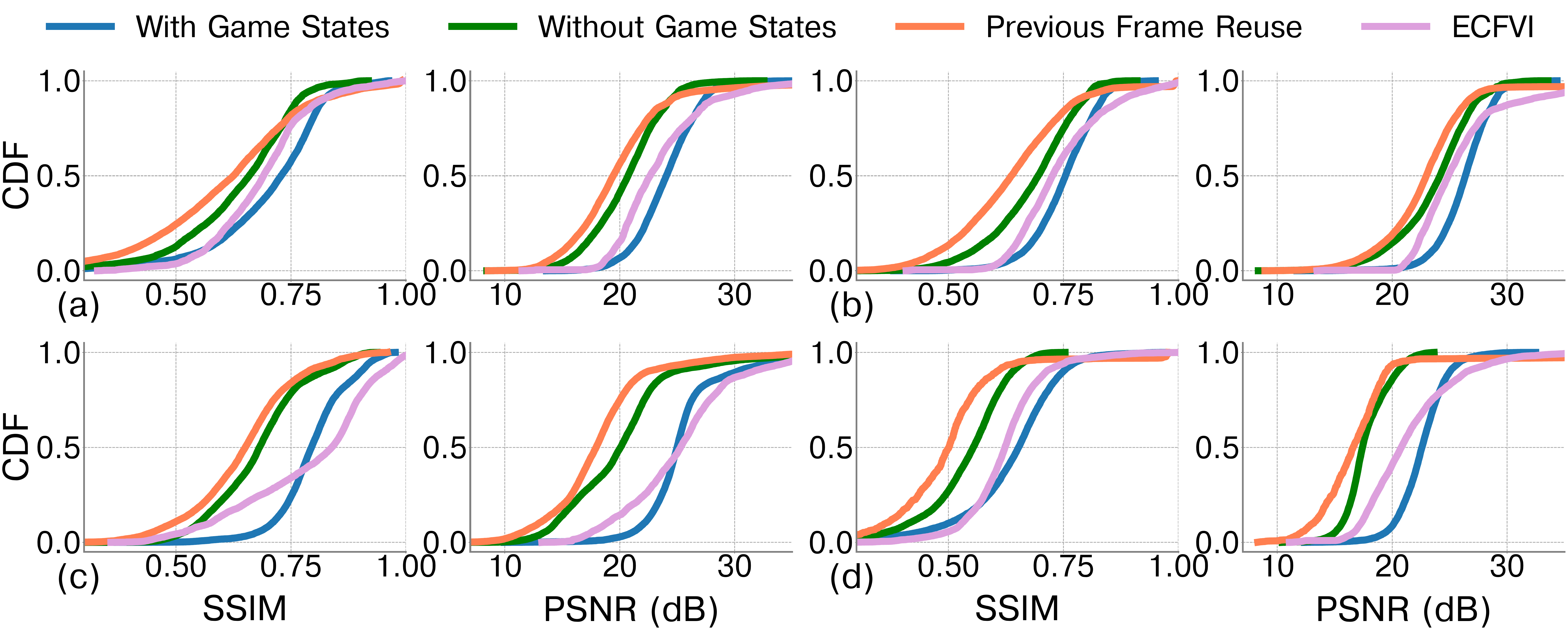}
    % \vspace{-10pt}
    \caption{Performance results for different games. (a) Viking Village. (b) Nature. (c) Corridor. (d) Viking Village+}
    % \vspace{-15pt}
    \label{fig:vary_game}
\end{figure}

\para{Impact of game states: } \revised{We compare the following schemes for different games to analyze the impact of game states: (i) our method with game states, (ii) our method without game states, (iii) reusing the last frame as the recovered frame, \revision{and (iv) ECFVI~\cite{kang2022error}, a state-of-the-art flow-guided video inpainting model}. Note that here we assume the previous frames are received correctly. In (ii), we use the last two consecutive frames to replace the game states input in Figure~\ref{fig:arch} because they are usually used to get an optical flow in traditional video recovery algorithms. \revision{In (iv), ECFVI is limited to only restoring a 240p video frame. To enhance this, we introduce additional convolutional layers and incorporate a pixel shuffle layer to upscale from 240p to 1080p. This is trained using our dataset and evaluated at the same epoch as our model.} 

As shown in Figure~\ref{fig:vary_game}, for Viking Village the improvements of (i) over (ii), (iii), and (iv) are 0.06, 0.1, and \revision{0.012} in SSIM, respectively;  3.5dB, 4dB, and \revision{0.7dB} in PSNR, respectively; for Nature the corresponding improvements are 0.07, 0.12, and \revision{0.008} in SSIM, respectively; 2.6dB, 3.1dB, and \revision{0.4dB} in PSNR, respectively; for Corridor the corresponding improvements are 0.11, 0.15, and \revision{0.006} in SSIM, respectively; 5.6dB, 7.3dB, and \revision{0.3dB} in PSNR, respectively; for Viking Village+ the corresponding improvements are 0.09, 0.14, and \revision{0.016} in SSIM, respectively, and 4.9dB, 5.6dB, and \revision{0.9dB} in PSNR, respectively. 

We make the following observations. First, using game states significantly improves video quality over recovery without game states. Second, reusing previous frames only performs slightly worse than recovery without game states due to little change in consecutive video frames. \revision{Third, our light-weight video recovery model based on game states yield even better results than ECFVI, a much bigger model. We implement ECFVI using the same  CoreML optimization and evaluate it on an Apple MacBook Air, as its high memory requirements prevent it from being deployed on an iPhone12. Therefore, we compare our implementation of ECFVI with our approach on Macbook Air and find ECFVI takes 6s to make an inference while our model takes only 18ms. This demonstrates the efficiency of our model and the benefit of 
incorporating game states.}

\begin{table}[t!]
  \centering
  \resizebox{0.65\columnwidth}{!}{%
  \begin{tabular}{c|cc|cc|cc}
    \toprule
    & \multicolumn{2}{c|}{Viking Village} & \multicolumn{2}{c|}{Nature} & \multicolumn{2}{c}{Corridor} \\ \hline
    Recovery Type & SSIM & PSNR & SSIM & PSNR & SSIM & PSNR \\
    \midrule
    Partial & 0.748 & 25.93 dB & 0.798 & 28.74 dB & 0.842 & 27.72 dB \\ \hline
    Complete & 0.692 & 24.07 dB & 0.740 & 25.58 dB & 0.771 & 24.94 dB \\
    \bottomrule
  \end{tabular}
  }
  \vspace{10pt}
  \caption{Performance results for partially corrupted and completely lost frames}
  % \vspace{-10pt}
  \label{tab:pred_recovery}
  % \vspace{-11pt}
\end{table}

\begin{figure}[t!]
    \centering
    \includegraphics[width=1\columnwidth]
    {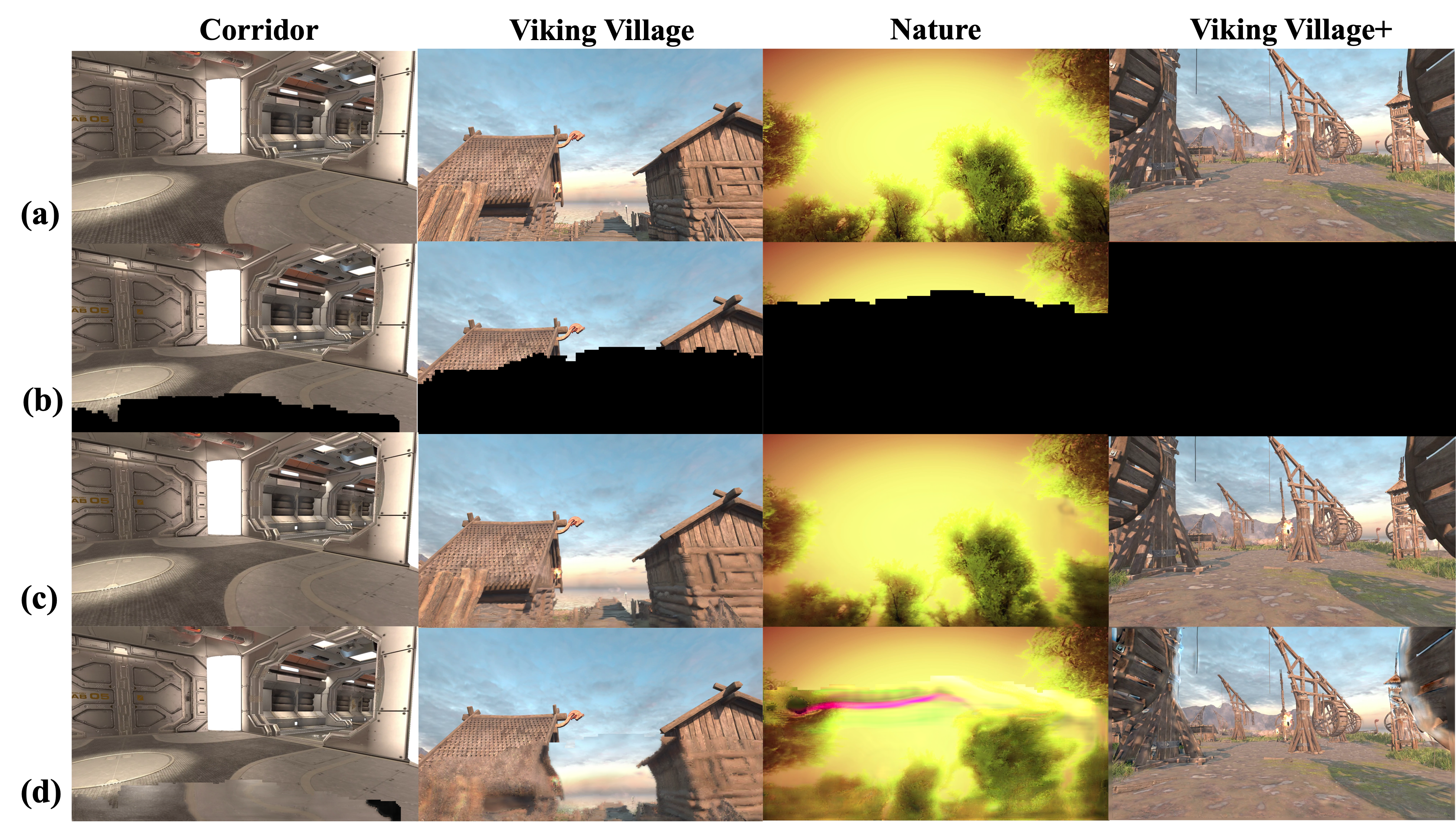}
    % \vspace{-20pt}
    \caption{Visualizations analysis of our recovery model using different masks for different games. The first three columns are the results on the iPhone 12, and the last column is the result on the laptop. (a) Ground truth. (b) Partial frame extracted from the decoded frame using mask. (c) Recovered frame. (d) Recovered frame without game states.}
    % \vspace{-15pt}
    \label{fig:CL}
\end{figure}

Table~\ref{tab:pred_recovery} shows the recovered video quality of partially corrupted frames and completely lost frames. Since the former has a portion of the original frame, its recovered video quality is higher than that of the latter. \revised{Figure~\ref{fig:CL} shows the visualization of our recovery model using different masks for different games. It indicates that our method can effectively recover a partially corrupted frame and stitch it with the correct portion together for an accurate result. In comparison, it is difficult to predict the optical flow without game states such that the recovered frames are blurred and corrupted.}

\comment{
\begin{figure*}[t!]
    \centering
    \includegraphics[width=1.7\columnwidth]{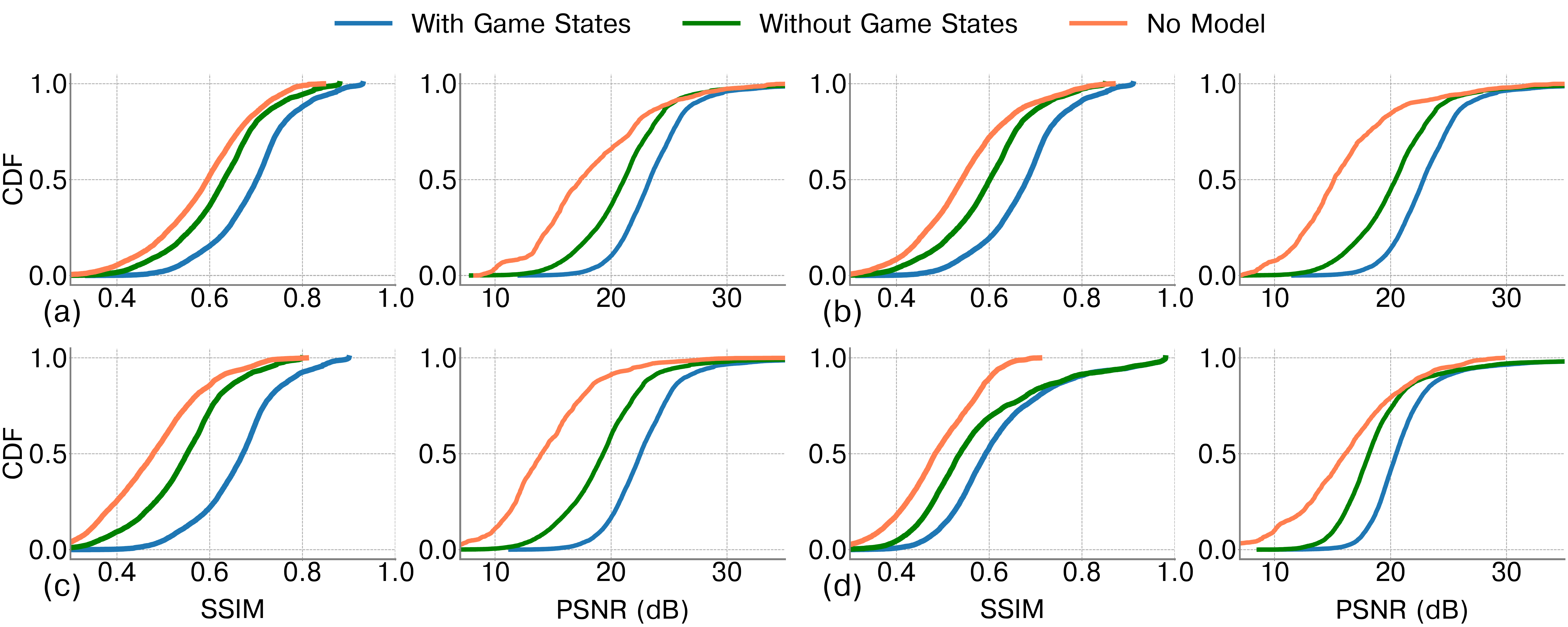}
    \vspace{-10pt}
    \caption{Results using different network traces. Viking Village: (a) 1\% loss rate (b) 3\% loss rate (c) 5\% loss rate. Viking Village+: (d) 1\% loss rate.}
    % \vspace{-10pt}
    \label{fig:vary_game_network}
\end{figure*}
}

\begin{figure}[ht]
    \centering
    % \vspace{-7pt}
    \includegraphics[width=0.8\columnwidth]{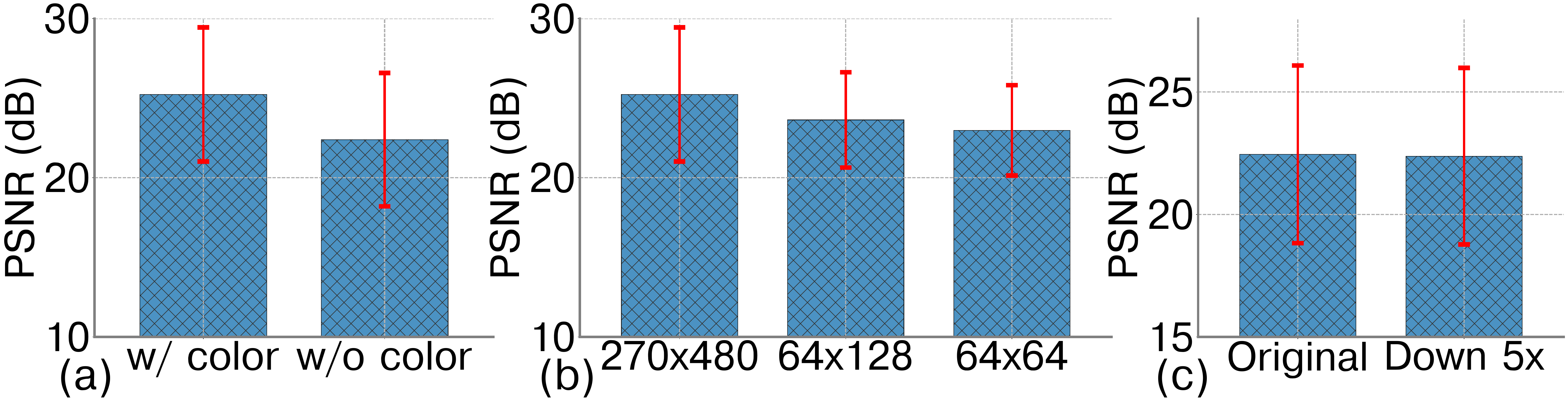}
    % \vspace{-10pt}
    \caption{Performance results for game states. (a) w/ and w/o color. (b) different sizes. (c) different downsampling ratio.}
    \vspace{-5pt}
    \label{fig:ablation}
\end{figure}

% \vspace{-2pt}
\para{Impact of game state representation: } As shown in Figure~\ref{fig:ablation}, we study the impact of different game state representations in three dimensions: color, size, and downsampling ratio. \newrevised{For brevity, we only show PSNR results.} We observe the followings: 1) After adding color, video quality has been improved by 2.9dB in PSNR as objects with different colors are better distinguished;
% 0.04 in SSIM and 
2) We compare three sizes of game states: $270\times480$, $64\times128$, and $64\times64$. While larger sizes yield performance improvement, smaller game states can still yield acceptable performance. For example, using a $64\times64$ image to warp the previous frame achieves 
% 0.72 SSIM and 
23 dB PSNR. 3) Downsampling the vertices corresponding to the game objects by a factor of 5 has minimal effect on accuracy, while saving 66.4\% and 67.5\% of the computation time on the iPhone 12 and laptop.

% \vspace{-2pt}

\begin{figure}[t!]
    \centering
    \includegraphics[width=1\columnwidth]
    {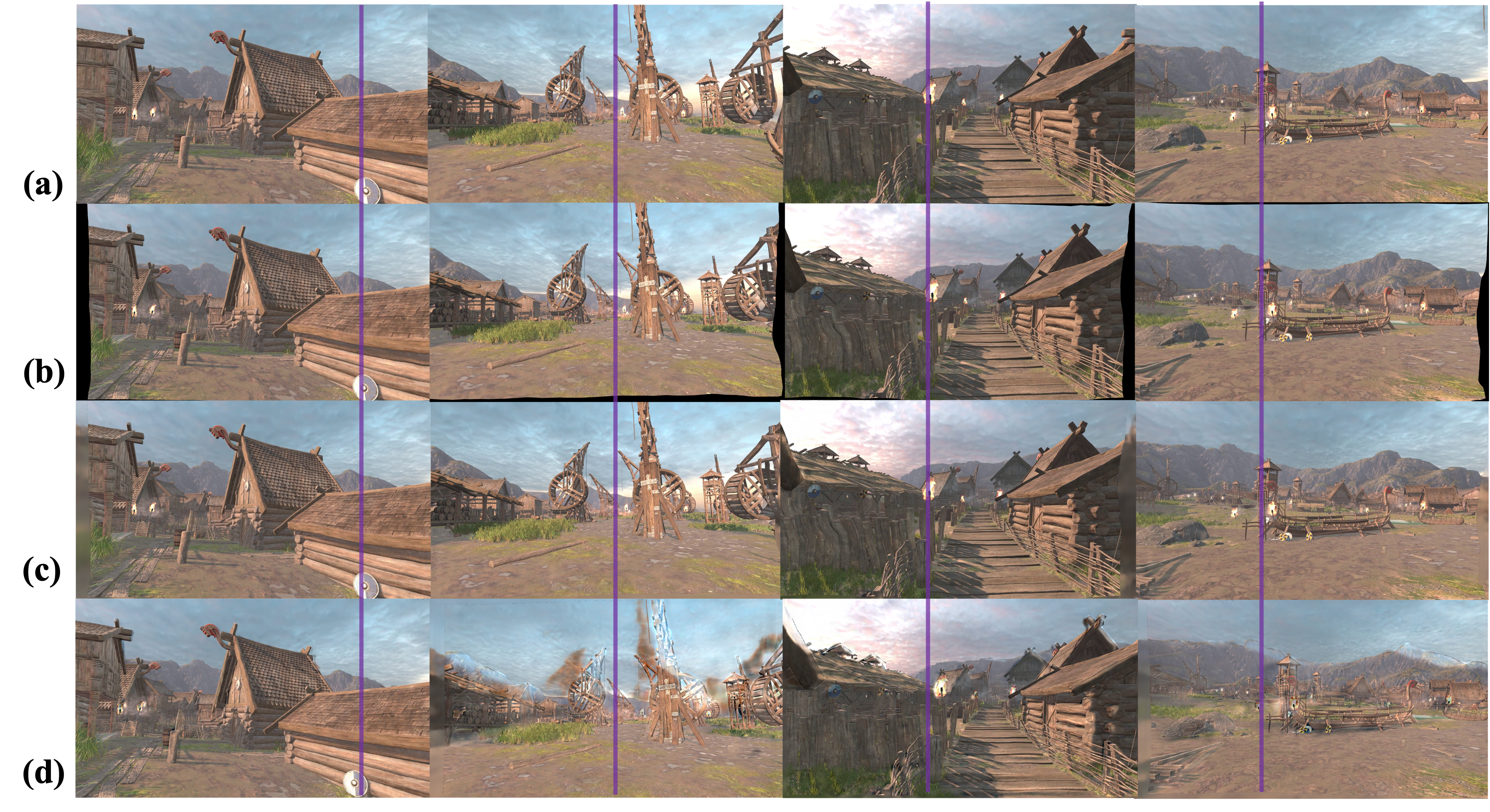}
    % \vspace{-20pt}
    \caption{Visualizations analysis of enhancement and inpainting. Here shows the results of Viking Village+ on the laptop. A purple auxiliary line is added to reflect the alignment to the ground truth. We make a whole frame prediction. (a) Ground truth. (b) Warped result. (c) Recovered frame. (d) Recovered frame without game states.}
    \label{fig:regen_vis}
    % \vspace{-15pt}
\end{figure}

\para{Enhancement and Generation:} \revised{As shown in Figure~\ref{fig:regen_vis}, row (b) shows the limitation of warping. Warping can only move pixels to achieve alignment but cannot generate new objects. For areas where no reference can be found from the previous frame, black borders appear. Row (c) shows that the black edge is reasonably filled in due to the inpainting module. The inpainting module restores the black area after warping, where the game states additionally serve as a hint for the generation because it contains the position information of all objects in the frame. Row (d) shows that the inaccurate flow causes the warping to be poorly aligned when no game states are applied. However, the enhancement module fixes the warped result such that the color of the inaccurately positioned object is suppressed and some correct colors are regenerated in the aligned area. Nevertheless, regeneration from scratch results in blurred and inaccurate predictions. Therefore, game states help to obtain a high-quality optical flow which is important to align features.}

\begin{figure}[t!]
    \centering
    \includegraphics[width=0.9\columnwidth]{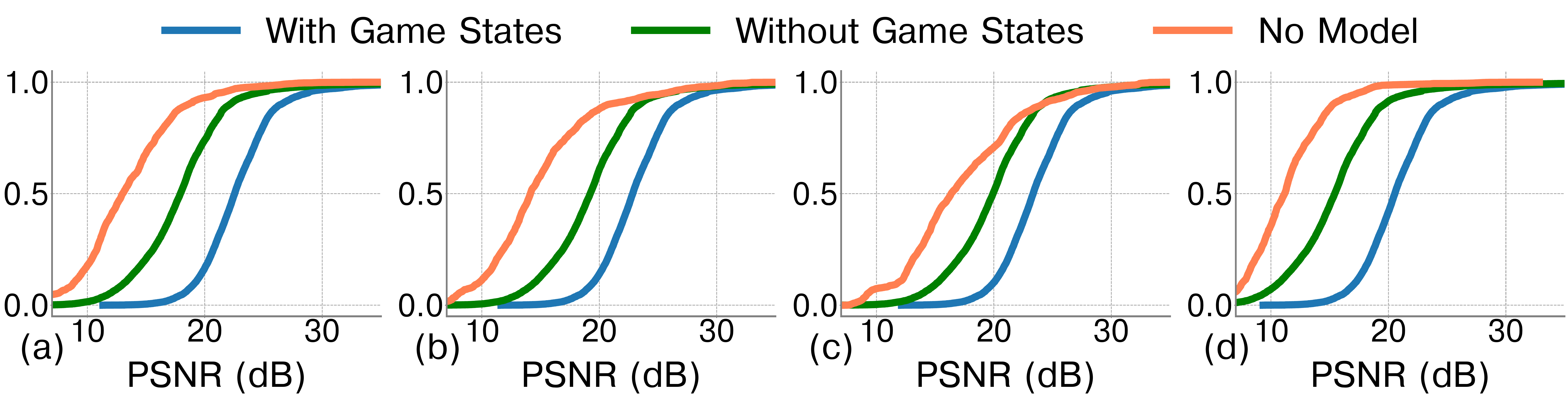}
    % \vspace{-10pt}
    \caption{Performance results for different network traces. (a) 4G. (b) 5G. (c) WiFi. (d) LEO.}
    % \vspace{-15pt}
    \label{fig:vary_game_network}
\end{figure}

% \vspace{-2pt}
\para{Impact of different network conditions: } \revised{We further evaluate different network traces to analyze the impact of realistic network losses. We recover corrupted frames  using  previously rendered or recovered ones. In this case, the recovered frame will be fed to the model as an input for the next frame recovery. For better comparison, we compare only the recovered frames. We consider three schemes: (i) recovery using game states, (ii) recovery without game states, and (iii) without our recovery model. Note that in (iii), our video decoder will reuse previous frames if the current frame is completely lost.} \newrevised{Figure~\ref{fig:vary_game_network} shows the recovery performance for Viking Village under 4G, 5G, WiFi, and LEO networks. 
% We separate driving scenarios from 4G and 5G traces because they have a larger throughput variation and larger loss rate than static and walking scenarios. 
(i) out-performs (ii) by 4.8dB, 3.9dB, 3.7dB, and 5.4dB, respectively. (i) also improves (iii) by 9.1dB, 8.2dB, 6.4dB, and 9.7dB, respectively. As expected, (i) performs the best by using game states. The gain of (i) over (ii) is slightly improved compared with that in Figure~\ref{fig:vary_game}(a) because (ii) uses the last two consecutive frames for recovery, so the recovery error will be more likely to accumulate. (iii) performs the worst because it cannot recover corrupted frames and its performance degrades fast under bursty packet losses. As shown in Table~\ref{tab:network_traces}, the packet loss rate is 3.6\%, 2.5\%, 1.1\%, and 9.4\% for 4G, 5G, WiFi, and LEO traces, respectively. We observe that the overall PSNR degrades as the loss rate increases across different traces. Recovering using game states achieves a larger gain over the others as the loss rate increases due to fast error accumulation without game states.} 

\revision{
\para{Impact of scheduling recovery: } We let the client start extracting game states for a video frame if the previous frame is lost or misses its deadline. In this way, video recovery is executed using the most recently rendered frame. So, if a frame A is the first missed frame, the subsequent frame will be recovered using A's predecessor. The game state extraction for both A's successor and predecessor occur simultaneously since we retain the user's previous inputs in cache. On the iPhone12, this parallel extraction results in a $0.8ms$ increase in game state extraction time, which is negligible.}
\eurosys{We evaluate the impact of this method over different networks, especially during long burst of frame losses. Figure~\ref{fig:psnr_reduction} plots the PSNR reduction along with the average number of consecutive corrupted frames for different networks. The average number of consecutive corrupted frames is 113, 89, 62, and 150 for 4G, 5G, WiFi, and LEO networks, respectively. Correspondingly, the average PSNR is reduced by only 0.17dB, 0.13dB, 0.06dB, and 0.26dB. The minimal drop in quality is largely attributed to the usage of game states, which offer significant details about the corrupted frame for recovery. It also benefits from our enhancement and inpainting module, which effectively corrects inaccurate warped pixels. Note that if a portion of a video frame is received correctly, we can use the received pixels to replace the predicted pixels.}

\begin{table}[t!]
  \centering
  \resizebox{0.9\columnwidth}{!}{%
  \begin{tabular}{c|cc|cc|cc|cc}
    \toprule
    Network Traces & \multicolumn{2}{c|}{4G} & \multicolumn{2}{c|}{5G} & \multicolumn{2}{c}{WiFi} & \multicolumn{2}{c}{LEO} \\ \hline
    Recovery Type & Our & 40\% FEC & Our & 35\% FEC & Our & 25\% FEC & Our & 70\% FEC \\
    \midrule
    Avg. pixel loss rate (\%) & 71 & 28 & 63 & 22 & 57 & 12 & 83 & 41 \\ \hline
    Avg. PSNR (dB) & 31.65 & 29.01 & 33.53 & 31.58 & 37.51 & 36.8 & 27.95 & 23.26 \\
    \bottomrule
  \end{tabular}
  }
  \vspace{10pt}
  \caption{Comparison with forward error correction (FEC)}
  % \vspace{-10pt}
  \label{tab:compare_fec}
  \vspace{-10pt}
\end{table}

\para{Comparison with FEC: }\newrevised{Table~\ref{tab:compare_fec} compares average pixel loss rate and PSNR between our approach and oracle FEC. which knows the network loss rates in advance.  To prevent packet loss, 40\%, 35\%, 25\%, and 70\% FEC are used in 4G, 5G, WiFi, and LEO network traces, respectively. However, due to extra FEC transmissions, some frames may not be delivered within 33ms deadline, resulting in frame losses. In comparison, our approach achieves higher video quality over FEC: it improves PSNR by 0.7-4.7dB and incurs less transmission overhead. Moreover, in practice it is challenging to realize the oracle FEC due to unpredictable network loss rate, and the performance gap between our approach and practical FEC is likely to be even larger.}

\begin{figure*}
\centering
\begin{minipage}[ht]{1\columnwidth}
\begin{minipage}{0.5\columnwidth}
    \centering
    \vspace{-2pt}
    \includegraphics[width=0.8\columnwidth]{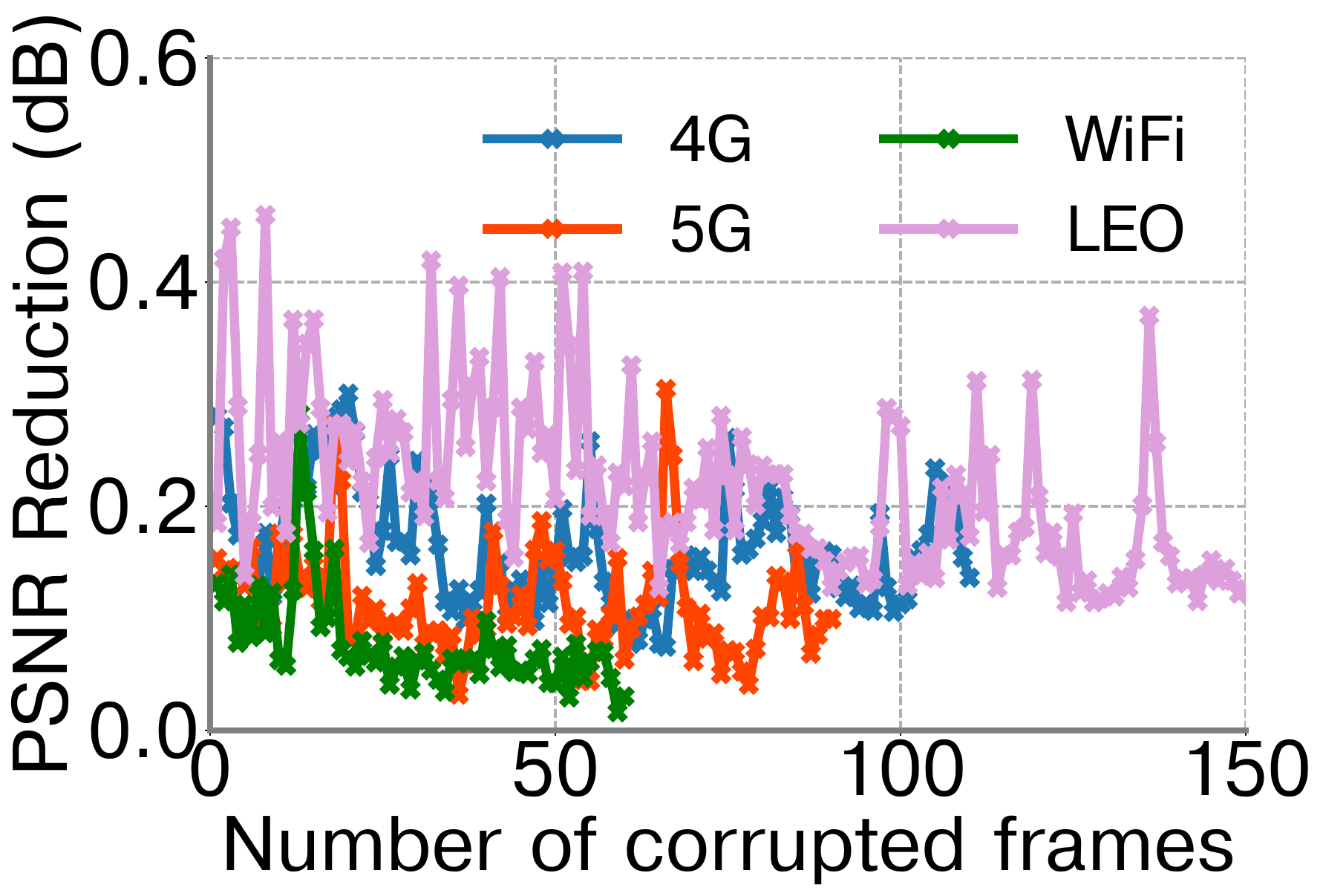}
    % \vspace{-10pt}
    \caption{PSNR reduction with the number of consecutive corrupted frames.}
    \label{fig:psnr_reduction}
    \vspace{-10pt}
\end{minipage}
\hspace{5pt}
\begin{minipage}{0.5\columnwidth}
    \centering  
    \includegraphics[width=0.8\columnwidth]{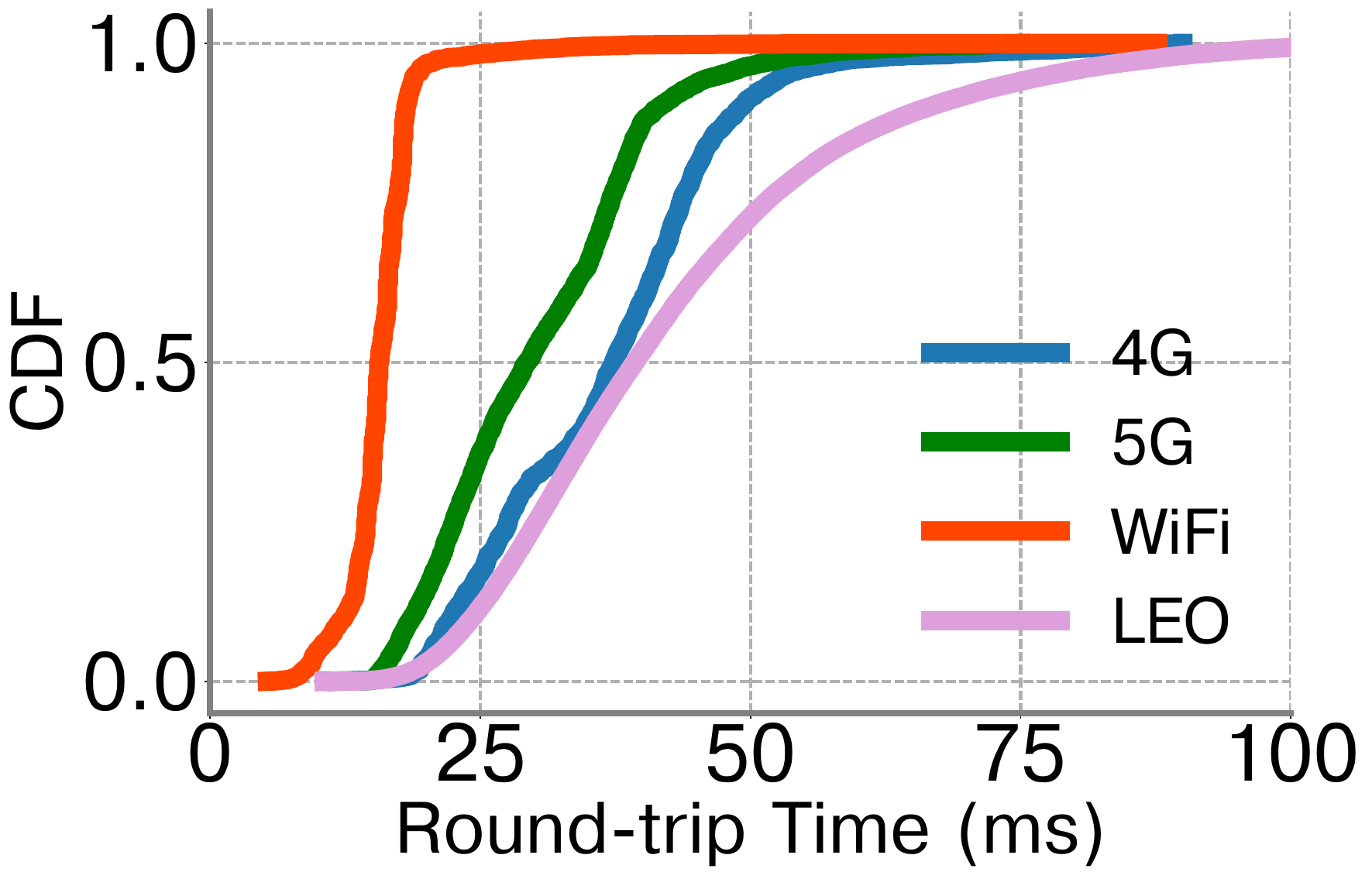}
    % \vspace{-10pt}
    \caption{\revision{CDF of Round-trip Time (RTT) for different networks.}}
    \label{fig:rtt_cdf}
    \vspace{-10pt} 
\end{minipage}
\end{minipage}
\end{figure*}

\revision {
\subsection{End-to-end latency} 
We evaluate the end-to-end latency when playing Viking Village, Nature, Corridor, and Viking Village+. Our game server operates on a desktop equipped with an AMD 12-Core Ryzen 9 5900 CPU, Nvidia GeForce RTX 3080 graphics card, and a 2TB SSD. Meanwhile, the client operates on an iPhone12, connected via 4G, 5G, WiFi, and LEO networks.} \eurosys{We first evaluate the latency without the need of recovery and the components are as follows:}

\begin{equation}
\begin{aligned}
    \underset{(30.8-66.6ms)}{T_{non-recovery}} = \underset{(15-42ms)}{RTT} + \underset{(2.9-8ms)}{T_{server\_render}} + \underset{(4.3-5ms)}{T_{server\_encode}} + \underset{(1.6-4ms)}{T_{tx}} + \underset{(7-7.6ms)}{T_{client\_decode}}
\end{aligned}
\label{eqn:non-recovery}
\end{equation}

\begin{figure}[t]   
    \centering   
    \label{tab:breakdown}  
    \resizebox{0.42\columnwidth}{!}{%  
      \begin{tabular}{cccc}  
        \toprule  
        \multirow{2}{*}{\centering Game Name} & Server & Server & Client \\
        & Render & Encode & Decode \\ 
        \midrule  
        Viking Village & 3.6ms & 4.5ms & 7.2ms \\  
        Nature & 2.9ms & 4.3ms & 7ms \\  
        Corridor & 3.2ms & 4.8ms & 7.4ms \\ 
        Viking Village+ & 8ms & 5ms & 7.6ms \\ 
        \bottomrule  
      \end{tabular}  
    }  
    \caption{\revision{Latency breakdown of server and client's processing time when running different games}} 
    % \vspace{-10pt}   
\end{figure}  

\revision {
\begin{itemize}
    \item $RTT$: This is a Round-trip time (RTT) involved in sending a user’s input and receiving the resulting video frame from the server (excluding the transmission time, which is accounted for in $T_{tx}$). Figure~\ref{fig:rtt_cdf} illustrates the CDF of the observed RTTs. The average RTT is $36ms$, $30ms$, $15ms$, and $42ms$ for 4G, 5G, WiFi, and LEO networks, respectively.
    \item $T_{server\_render}$ and $T_{server\_encode}$: Upon receiving a user input from the client, the server processes and encodes the video frame using the H.264 codec. Table ~\ref{tab:breakdown} illustrates that the server's rendering time varies based on the game's complexity, ranging from $2.9-8ms$ for the games we evaluated. For a fixed output resolution and encoding parameters, the encoding process takes $4.3-5ms$, which has minimal fluctuation.
    \item $T_{tx}$: The duration required to transmit a frame from the game server to the client is influenced by the network bandwidth. Transmitting a 1080p video frame takes approximately $3.6ms$ over 4G, $1.9ms$ over 5G, $1.6ms$ over WiFi, and $4ms$ over LEO networks.
    \item $T_{client\_decode}$: For the games we evaluated, as indicated in Table~\ref{tab:breakdown}, the client on the iPhone12 processes and decodes an incoming video frame in $7-7.6ms$.
\end{itemize}

The end-to-end latency may fluctuate depending on the network condition, system load, and the complexity of the game. When a video frame does not arrive before the deadline either due to a fluctuating delay or a lost frame, our system prompts the client to extract game state and execute video recovery for the next frame. }\eurosys{As explained in Sec.~\ref{sec:system}, the end-to-end latency with recovery is as follows:}

\begin{equation}
\begin{aligned}
    \underset{(<=80ms)}{T_{recovery}} = max(min(\underset{(58ms)}{Timeout}, \underset{(30.8-66.6ms)}{T_{non-recovery}}), \underset{(7ms)}{T_{gs}}) + \underset{(22ms)}{T_{inference}}
\end{aligned}
\label{eqn:recovery}
\end{equation}

\eurosys{With an 80ms deadline budget and a model inference time of 22ms, the timeout can be adjusted to 58ms. The timer begins as soon as the user input is generated. Video recovery will be initiated either when the timer runs out or when a corrupted frame arrives. The game state extraction can be done in parallel with all steps except the model inference. It's important to highlight that if pixels are received post-timeout but prior to the deadline, they can replace the recovered pixels. We also have the flexibility to vary the timeout for different latency budgets. The mechanism ensures our system's capability to achieve real-time recovery for any missing or corrupted frame. Successfully recovered frames are stored locally and used for potential future recovery.}

\yifan{
\subsection{Resource Usage}
In Sec. 4.3, the additional memory overhead for Viking Village and Viking Village+ is 200MB and 630MB, respectively, well below smartphones' storage. Modern smartphones have a shared memory between the CPU and GPU. As the memory capacity of mobile devices continues to increase (e.g., 8GB, 16GB, 32GB),  our memory requirement is quite affordable. 
With no frame loss, the server renders and transmits all video frames, similar to traditional approaches. In this scenario, iPhone 12's CPU utilization is 35\% and energy consumption is 0.05J per frame. Under 20\% frame losses, the corresponding numbers are 41\% and 0.06J per frame, and under 100\% frame losses, they are 74\% and 0.08J per frame. Consequently, in the worst case when every frame needs recovery, the battery life decreases from 10.5 hours to 6.6 hours. 
% In practice, since the video frame loss rate is generally below XXX, the battery life reduces by within XXX. 
}

% \vspace{5pt}
\section{Conclusion}
\label{sec:conclusion}
% \vspace{-1pt}

\revised{In this paper, we develop the first deep learning system for cloud gaming video recovery on mobile devices. Our system efficiently extracts and utilizes game states, generates a mask from H.264 video decoder to indicate which portions of frames need recovery, and develops a novel video recovery model to recover completely lost or partially corrupted frames to cope with server overload, network congestion, and losses in cloud gaming. Our extensive evaluation under diverse network conditions shows the effectiveness of our approach on mobile devices. We hope our work will inspire follow-up work to further enhance cloud gaming performance in challenging network environments.
}

\bibliographystyle{abbrv}
\bibliography{video,video2}

\end{document}